\newlength\figwidth \figwidth=3.3 in
 \documentclass[aps,pra,twocolumn,floatfix,amsmath,amssymb,showkeys,letterpaper]{revtex4}

\newcommand{\be}{\begin{equation}}
\newcommand{\ee}{\end{equation}}

\newcommand{\sgn}{\, \text{sgn}}

\newcommand{\bn}{\begin{eqnarray}}
\newcommand{\en}{\end{eqnarray}}

\newcommand{\bsub}{\begin{subequations}}
\newcommand{\esub}{\end{subequations}}

\usepackage{graphicx}
\usepackage{amsmath}
\usepackage{amsfonts}
\usepackage{hyperref}
\usepackage{bm}
\usepackage{color}
\definecolor{color1}{rgb}{0.4,0.0,0.8}
\definecolor{color2}{rgb}{0.6,0.0,0.0}
\usepackage{soul}

\begin{document}
\title{Domain walls with alternating magnetic order in a model with dipolar coupling}
\author{G. M. Wysin}
\email{wysin@k-state.edu}
\homepage{http://www.phys.ksu.edu/personal/wysin}
\affiliation{Department of Physics, Kansas State University, Manhattan, KS 66506-2601}

\date{August 28, 2026}
\begin{abstract}
A model for a one-dimensional chain of elongated nano-scale magnetic islands with dipole interactions is
analyzed here for the properties of its static domain walls with site-by-site alternating order. 
The anisotropic magnetic islands on a nonmagnetic substrate have their longer axes oriented transverse 
($y$-direction) to the chain direction ($x$-direction), in a transverse applied magnetic field. 
The islands' magnetic dipoles $\bm{\mu}_n$ are represented as macrospins of fixed length $\mu$.
%
The nearest-neighbor (NN) dipole interactions drive transverse alternating order, allowing for doubly-degenerate, 
uniform, static, $y$-alternating states, where the dipoles alternately point transverse to the chain direction, 
as in $\bm{\mu}_n = \pm(-1)^n \mu \hat{y}$. 
Assuming only NN interactions, the domain walls connecting these two alternating states are found with 
numerical relaxation simulations and analyzed in a two-sublattice continuum theory.
As the uniaxial anisotropy constant $K_1$ descends from larger values until it closely approaches the NN dipolar 
coupling constant $D$, the domain wall width grows indefinitely, and the large-$x$ continuum solutions closely 
approach the lattice numerical solutions.  
The applied field produces a very slight canting of the dipoles towards the field, maximum in the center of the 
domain wall.  
The dipoles on the two sublattices are found to rotate in {\em opposite senses} as one scans along 
the lattice, resulting in a large longitudinal magnetic moment of the domain wall. 
A much smaller transverse magnetic moment has a topological contribution that depends on whether the chain 
length is odd or even.
\end{abstract}

\maketitle

\section{Introduction: Multi-stable magnetic island chains}
Nanoscale engineered magnetic arrays such as two-dimensional (2D) spin ices of many 
geometries \cite{Wang06,Nisoli13,skjaervo19} and one-dimensional (1D) chains of thin magnetic 
islands \cite{Ostman18,Cisternas21,Anand21}, existing on nonmagnetic substrates, are systems whose properties might 
be tunable for desired functions.  Both of these general groups of systems involve islands whose magnetic dipoles 
$\bm{\mu}_n$ interact via dipolar interactions that compete with island-shape anisotropies and with an applied field.  

These types of models have uniform metastable states, often existing in symmetry-related pairs, differing 
only in some quantum number related to that physical symmetry.  The symmetry may invert or rotate some 
spin component(s) to convert one of the paired uniform states into the other, without a change in energy
while conserving some component of the total magnetic moment. With the existence of such pairs, 
domain walls (DWs) exist that connect them over some distance. I avoid referring to them as solitons. This
work does not establish any solitary scattering properties for the solutions being found. The analysis is 
simplest in 1D chains. 

\begin{widetext}

\begin{figure}
\includegraphics[width=2\figwidth,angle=0]{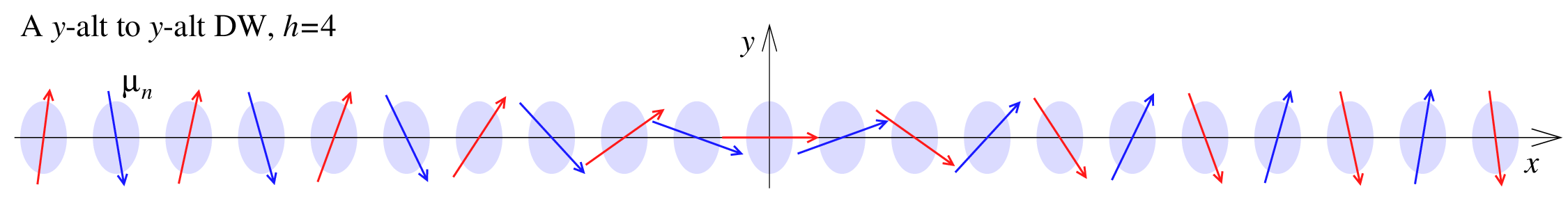}
\caption{\label{oval-islands-yaltdw} A static $y$-alt$^2$ DW centered in a system of 21 islands, with a half-width $h=4$ 
in lattice constants, embedded between the two possible $y$-alt states.  The red/blue arrows represent the island dipoles on 
odd/even sublattices, respectively. The sublattices rotate through a 90$^{\circ}$ angle in opposite senses, 
while remaining in the $xy$-plane, as one scans along the chain. At the center, the dipoles can point towards $+x$ 
(as shown) or towards $-x$ by reversing the rotation senses on both sublattices.}
\end{figure}

\end{widetext}

Here I consider a model \cite{Wysin24} of a 1D chain (along $x$) of islands whose longer axes are oriented transverse (along $y$) 
to the chain, without and with a magnetic field applied transversely (along $+y$), and a small thickness along $z$
perpendicular to the substrate.  Without an applied field, there are three stable or metastable uniform states: 
\begin{enumerate}
\item $x$-parallel, where the island dipoles all point along $+x$ or $-x$. 
\item $y$-parallel (or $y$-par for short), where the island dipoles all point along $+y$ or $-y$. 
\item $y$-alternating (or $y$-alt for short), where the island dipoles alternate between $+y$ and $-y$ directions.  
\end{enumerate}
The two $y$-alt states are distinguished by either having even/odd dipoles along $+y/-y$ or along $-y/+y$, respectively,
see Fig.\ \ref{oval-islands-yaltdw}.
Note that a uniform $y$-alt configuration minimizes the nearest neighbor (NN) dipolar energy, while uniform 
$y$-parallel maximizes it.  

When a weak transverse magnetic field is applied, the $y$-par and $y$-alt configurations are unchanged, while
the dipoles in $x$-parallel states tilt away from $\pm x$ towards $+y$, resulting in oblique states (dipoles
at an oblique angle to $\pm x$).  The degeneracy of the two oblique states is obvious, and one is transformed 
to the other by reversing the $x$-components of the dipoles.  A DW can form between these two oblique
states, and the structure and creation energies of those oblique-to-oblique DWs were analyzed in recent
work \cite{Wysin26}. It was found that an approximate continuum limit for the angles of the dipoles in the $xy$-plane
leads to a scalar $\varphi^4$ equation, that is adequate to describe the oblique-to-oblique DWs.

Numerical simulations \cite{Wysin26} also indicated the existence of stable DWs from oblique to $y$-alt when their energies
are close, and between the two degenerate $y$-alt states.  It is difficult to imagine a continuum theory
going smoothly from a nearly aligned oblique state to the anti-aligned $y$-alt state. In contrast, a continuum theory
for a configuration connecting the two $y$-alt states is realizable and developed in this work. 

The goal of this report is to develop a continuum theory for the description of the $y$-alt to $y$-alt DWs,
which are referred to as $y$-alt$^2$ DWs.  An example is shown in Fig.\ \ref{oval-islands-yaltdw}.
That includes finding the variation of the dipoles from one side of the DW to the other, as well as its
net magnetic moment and creation energy.  This is done for a limited range of shape anisotropy relative
to the NN dipolar coupling.  The theory is restricted to finding static DWs with NN dipolar coupling; 
long-range interactions are likely to shift some results.  Numerical simulations of discrete islands
in a chain are used to guide and verify the continuum limit and indicate where it fails when
the DW becomes too narrow.

\subsection{Approaches to find $y$-alt$^2$ DW solutions}
In order to verify the static $y$-alt$^2$ DW solutions, I present three methods for their determination.

The first method is numerical relaxation simulations, using the scheme in Ref.\ \cite{Wysin26}, where the 
macrospins are iteratively pointed in the direction of the effective magnetic field that acts on them.  This results
in zero torque on each spin, together with a local minimum of energy, hence a stable static DW. 

The second method is to find continuum equilibrium equations for the dipole directions by making approximate 
expansions of the discrete equations of motion for the island dipoles. The continuum theory is developed similar 
to the two-sublattice approach used for antiferromagnetic chains \cite{Mik80,FlugMik83,GalIv18}.  
The equilibrium equations so obtained have approximate asymptotic solutions far from the DW center that 
turn out to work surprisingly well even near the center of the DW.

In a third method, the continuum theory was checked by starting from a continuum expansion of the Hamiltonian, 
and using it for development of the equilibrium equations. This last method confirms those equilibrium equations, 
but only with careful consideration of up to third order small terms.

\section{The model for macrospins coupled by dipole interactions}
The islands are assumed to have individual magnetic dipole moments $\bm{\mu}_n = \mu{\bf S}_n$ of fixed magnitude
$\mu$, described by unit-length classical Heisenberg-like \cite{Heisenberg28,Jiles91} macrospins ${\bf S}_n$.  
They are each affected by a transverse uniaxial anisotropy 
of strength $K_1$ and an easy-plane anisotropy of strength $K_3$, as well as the applied magnetic field ${\bf B}=B \hat{y}$,
so that the Hamiltonian involving dipolar interactions among all $N$ dipoles on a chain is
\begin{align}
\label{Hn}
H = \sum_{n=1}^{N}  &  \Big\{ \sum_{k=1}^{R} \frac{D}{k^3}
\left[ {\bf S}_n\cdot {\bf S}_{n+k} -3({\bf S}_n\cdot \hat{x})({\bf S}_{n+k}\cdot \hat{x})\right]
\nonumber \\
& -K_1 \left(S_n^y\right)^2 +K_3 \left(S_n^z\right)^2 -\mu B S_n^y \Big\}.
\end{align}
The NN dipolar coupling constant is an energy,
\be
D \equiv \frac{\mu_0 \mu^2}{4\pi a^3},
\ee
where $\mu_0$ is the permeability of space and $a$ is the lattice constant or center-to-center separation of NN dipoles. 
The range of the dipole interactions is $R$.  Although dipole interactions are long range, in order to get initial 
progress, I set $R=1$ and confine most of the analysis to the NN model. This produces features similar to what happens
in the long-range model ($R\to\infty$), but at different coupling parameters. For a finite chain the interaction sum is 
cut off when $n+k > N$.  

For classical spin mechanics, a magnetic dipole is related to its angular momentum by 
$\bm{\mu}_n = \gamma{\bf L}_n = \mu {\bf S}_n$, where $\gamma$ is a gyromagnetic ratio.  
With unit magnitude ${\bf S}_n$ being dimensionless, the angular momentum of a dipole is $\mu/\gamma$. 
The rate of change of ${\bf L}_n$ is equal to the net torque on that dipole,
which gives the basic dynamic equation,
\be
\label{dyn1}
\frac{1}{\gamma} \frac{d\bm{\mu}_n}{dt} = \bm{\mu}_n\times {\bf B}_n 
= \bm{\mu}_n\times\left( -\frac{\partial H}{\partial \bm{\mu}_n} \right).
\ee
${\bf B}_n$ is the net effective magnetic field acting on a dipole, caused by dipolar and anisotropic
effects as well as the applied magnetic field. 

The classical equations of motion can be developed in terms of the Cartesian spin components, however, it is
more advantageous to use planar-spherical angular coordinates, representing the macrospin components as
\begin{align}
{\bf S}_{n} & = (S_n^x,\ S_n^y,\ S_n^z)
\nonumber \\
& = (\cos\theta_n\cos\phi_n,\ \cos\theta_n\sin\phi_n,\ \sin\theta_n).
\end{align}
For one dipole, the angle $\phi_n$ can be viewed as a coordinate, and $(\mu/\gamma) S_n^z=(\mu/\gamma)\sin\theta_n$ is the
momentum conjugate to that coordinate. Then the vector dynamic equation is equivalent to Hamiltonian dynamic
equations for the angles, 
\begin{align}
\label{HamEq}
\frac{d}{dt}\phi_n = \frac{\partial H}{\partial \left(\frac{\mu}{\gamma} \sin\theta_n\right)},
\qquad
\frac{d}{dt}\left(\tfrac{\mu}{\gamma}\sin\theta_n\right) = -\frac{\partial H}{\partial \phi_n}.
\end{align}
%
%

The system Hamiltonian limited to NN interactions is expressed using angular coordinates as
\begin{align}
\label{Hphitheta}
H & = \sum_{n=1}^{N} \Big\{ D \big[\sin\theta_n\sin\theta_{n+1}+\cos\theta_n \cos\theta_{n+1}
\\
& \times (-2\cos\phi_n\cos\phi_{n+1}+\sin\phi_n\sin\phi_{n+1})\big] 
\nonumber \\
& -K_1\cos^2\!\theta_n \sin^2\!\phi_n +K_3\sin^2\!\theta_n -\mu B \cos\theta_n\sin\phi_n \Big\}.
\nonumber
\end{align}
Then the discrete Hamiltonian dynamic equations for the in-plane angles are
\begin{align}
\label{dotphi}
\frac{\mu}{\gamma} & \frac{d}{dt}\phi_n  =\frac{\partial H}{\partial \sin\theta_n} =  
D \big[ \sin\theta_{n-1}+\sin\theta_{n+1}  \\ 
& \hspace{-2mm} +2\tan\theta_n\cos\phi_n (\cos\theta_{n-1}\cos\phi_{n-1}+\cos\theta_{n+1}\cos\phi_{n+1}) 
\nonumber \\
& -\tan\theta_n\sin\phi_n (\cos\theta_{n-1}\sin\phi_{n-1}+\cos\theta_{n+1}\sin\phi_{n+1}) \big]
\nonumber \\
& +2K_1 \sin\theta_n \sin^2\!\phi_n +2 K_3 \sin\theta_n +\mu B \tan\theta_n \sin\phi_n .
\nonumber 
\end{align}
For the out-of-plane angle, there results 
\begin{align}
\label{dottheta}
\frac{\mu}{\gamma} & \frac{d}{dt}\theta_n = -\sec\theta_n \frac{\partial H}{\partial \phi_n} \\
& = -D \big[ 2\sin\phi_n (\cos\theta_{n-1}\cos\phi_{n-1} +\cos\theta_{n+1} \cos\phi_{n+1})
\nonumber \\
& \hspace{3mm} +\cos\phi_n (\cos\theta_{n-1}\sin\phi_{n-1} +\cos\theta_{n+1} \sin\phi_{n+1}) \big]
\nonumber \\ 
& +2K_1 \cos\theta_n \sin\phi_n \cos\phi_n+\mu B \cos\phi_n.
\nonumber 
\end{align}
One can note that the dipolar terms in both dynamic equations contain NN angles surrounding the
central site $n$. For simplicity, define dimensionless anisotropies and applied field,
\be
k_1 \equiv \frac{K_1}{D}, \quad 
k_3 \equiv \frac{K_3}{D}, \quad 
b \equiv \frac{\mu B}{D}.
\ee
Energies will be found in units of $D$.

\section{Alternating DW structures found by numerical simulations}
%
\begin{figure}
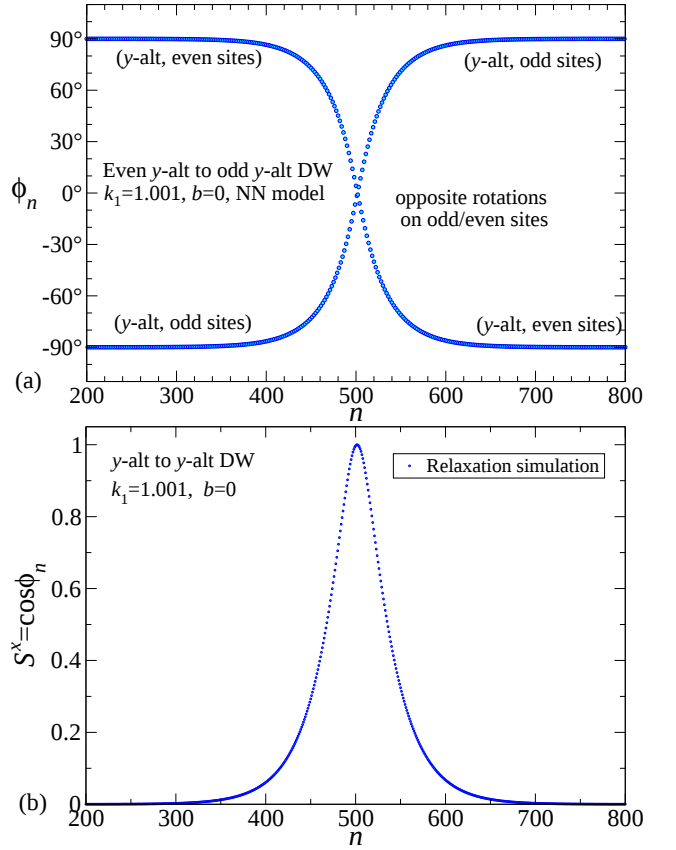

\includegraphics[width=\figwidth,angle=0]{yalt-yalt-nn-k1001-b0}
\includegraphics[width=\figwidth,angle=0]{sxN1001}
\caption{\label{phi1001} A $y$-alt to $y$-alt DW in the NN model from numerical relaxation simulations at 
$k_1=1.001$, $b=0$.   
(a) The in-plane angles, identified by odd/even sublattices.  
(b) The $x$-components of the macrospins. 
The view is zoomed in on the $N=1000$ system.  
The half-width $h=37a$ was estimated from the tangent line in $\phi_{n=\text{odd}}$ at the DW center. 
For $k_1$ farther above unity, the DW width is narrower.}
\end{figure}
It was noticed in previous work \cite{Wysin26} that DWs connecting the two $y$-alt uniform states exist, and 
that the spin structure within the DWs themselves has a site-by-site alternating component, similar to what 
occurs in solitons in antiferromagnetic (AFM) chains \cite{Mik80,FlugMik83}.  These DWs connect the 
even $y$-alt state ($\phi_n=+/- 90^{\circ}$ on even/odd sites) to the odd $y$-alt state 
($\phi_n=+/- 90^{\circ}$ on odd/even sites).  The $y$-alt states exist over a range of model parameters, 
especially constrained by $k_1>1$, and a limited value of $|b|$ that depends on $k_1$. 
The point $k_1=1$ is not included. Physically, at that point, the two $y$-alt states
could exist, but with infinite width.  For simplicity, we take $k_3=0$, although it would be present in real 
magnetic islands. Even the presence of any nonzero value of easy-axis anisotropy $k_1$ already nudges the dipoles 
to lie in the $xy$-plane, effectively acting as an easy-plane anisotropy. 

For $k_1=1.001$ and $k_3=b=0$, a $y$-alt$^2$ DW is displayed in Fig.\ \ref{phi1001}, where the odd 
$y$-alt state is on the left side of the system, connecting to the even $y$-alt state on the right side of the system.
This was obtained by starting with the two distinct $y$-alt states in each half of the system, and then
iteratively pointing the dipoles in the directions of their effective magnetic fields, ${\bf B}_n$. Further details
of the relaxation scheme are given in Ref.\ \onlinecite{Wysin26}. 
Fig.\ \ref{phi1001}a shows the in-plane angles $\phi_n$ on the discrete sites; the out-of-plane angles $\theta_n$ are zero
for any static configuration of the model.  It is significant that the dipoles make {\em opposing rotations} on the two 
sublattices as one moves along the chain. This is important when it comes to constructing a continuum limit for this problem. 
In Fig.\ \ref{phi1001}b, the macrospin components parallel to the chain direction are shown. It is notable that there 
is no distinction between the $x$-components on the two sublattices (for $b=0$ only). 
This type of DW is found only for $k_1 > 1$.

\section{Continuum equations derived from the discrete lattice equations}
\label{expand}
A continuum description of the DWs is helpful for predicting how their structures depend on the physical parameters 
$k_1,\ k_3$, and $b$. The intention is to find the continuum differential equations for describing the dipoles
along coordinate $x=na$.
To do this, the alternating structure must be accounted for. We observe that the dipole angles start at 
$\pm 90^{\circ}$ on the two sublattices (even vs.\ odd sites) on one side of the DW. They reduce in magnitude until 
reaching zero at the DW center, and then emerge to $\mp 90^{\circ}$ on the other side (Fig.\ \ref{oval-islands-yaltdw}).
They rotate in opposite senses as one passes through the DW along the chain (Fig.\ \ref{phi1001}a). 
Alternatively, they could have passed through $180^{\circ}$ at the DW center, i.e., forming a DW with the
opposite helicity.  The dipoles on the two sublattices maintain the same $x$-components through the DW, 
as in Fig.\ \ref{phi1001}b. The two sublattices' dipoles are parallel at the DW center, even for $b\ne 0$. 

These facts are incorporated by mapping the discrete site angles $\phi_n,\ \theta_n,$ into continuum {\em large} angles 
$\Phi(x)$ and $\Theta(x)$ and {\em small} angles $\varphi(x)$ and $\vartheta(x)$ according to the
{\em opposing rotations} assumption,
\begin{align}
\label{map}
\phi_n & \rightarrow \begin{cases} +\Phi(x)+\varphi(x), & n=\text{even}. \\ -\Phi(x)+\varphi(x), & n=\text{odd}. \end{cases}
\nonumber \\
\theta_n & \rightarrow \begin{cases} +\Theta(x)+\vartheta(x), & n=\text{even}. \\ -\Theta(x)+\vartheta(x), & n=\text{odd}. \end{cases}
\end{align}
One might consider the large angles as alternating or antiferromagnetic components and the small angles as
aligned or ferromagnetic components.
The small angles are generally considered one order smaller than the large angles. Later, $\varphi$ and $\vartheta$ are
shown to increase with the applied field $b$. At $b=0$, the small angles are zero. 
For opposing rotations at $b=0$, the two sublattices have the same $S^x$ but opposite $S^y$ and $S^z$.
That is in contrast to the {\em opposite directions} assumption generally used to describe DWs in 1D antiferromagnets,
where the two sublattices are approximately antiparallel in Mikeska's assumption 
\cite{Mik80,FlugMik83,Pires+89}.  However, 1D antiferromagnets apparently also support solitons with opposing
rotations on the sublattices, if one looks carefully enough \cite{Pires+89,Gerling+86}. 

\subsection{In-plane dynamic equations}
Now the discrete equations (\ref{dotphi}) are expanded into their equivalent continuum form, under the opposing rotations
assumption. Suppose a site $n$ is even.  In addition to Eq.\ (\ref{map}), we have for its first neighbors, both on
the odd sublattice, 
\begin{align}
\phi_{n+1} &= -\Phi(x+a)+\varphi(x+a), 
\nonumber \\
\phi_{n-1} &= -\Phi(x-a)+\varphi(x-a), \hskip 0.4cm  n=\text{even}.
\end{align}
Similar expressions apply to the out-of-plane angles. Then use Taylor expansions, keeping up to quadratic terms
in the small angles and in space derivatives of the large angles. Ignore spatial variations of the small angles.
\begin{align}
\Phi(x+a) & \approx \Phi(x)+a\Phi_x(x)+\tfrac{1}{2} a^2 \Phi_{xx}(x),
\nonumber \\
\Theta(x+a) & \approx \Theta(x)+a\Theta_x(x)+\tfrac{1}{2} a^2 \Theta_{xx}(x).
\end{align}
From here, I will write $\Phi$, $\Theta$ to indicate $\Phi(x)$, $\Theta(x)$, and the same for the derivatives.  The 
small angles do not require expansions. In addition, one needs expansions of various trigonometric functions, such as
\begin{align}
\label{cosa}
\cos\phi_{n+1} & \approx \cos[-\Phi(x+a)+\varphi(x+a)]
\nonumber \\
& \approx \cos[\varphi-\Phi-a\Phi_x-\tfrac{1}{2}a^2\Phi_{xx}]
\nonumber \\
& \approx \cos\Phi [1-\tfrac{1}{2}\left(\varphi^2+a^2\Phi_x^2-2a\varphi\Phi_x\right)]
\nonumber \\
& +\sin\Phi [\varphi-a\Phi_x-\tfrac{1}{2}a^2\Phi_{xx}].
\end{align} 
There is a corresponding expansion,
\begin{align}
\label{sina}
\sin\phi_{n+1} & \rightarrow \sin[-\Phi(x+a)+\varphi(x+a)] 
\nonumber \\
& \approx \sin[\varphi-\Phi-a\Phi_x-\tfrac{1}{2}a^2\Phi_{xx}]
\nonumber \\
& \approx  -\sin\Phi[1-\tfrac{1}{2}\left(\varphi^2+a^2\Phi_x^2-2a\varphi\Phi_x\right)]
\nonumber \\
& +\cos\Phi[\varphi-a\Phi_x-\tfrac{1}{2}a^2\Phi_{xx}].
\end{align}
There are similar expansions for the same functions of $\theta_{n+1}$. Further, these expressions give
the values at the opposite near neighbor $n-1$ by changing $a\to -a$.  Then, summing over left and
right nearest neighbors, the terms linear in $a$ cancel out, and one gets twice the even terms in $a$, 
such as the first dipole interaction term in Eq.\ (\ref{dotphi}),  
\begin{align}
d_1 & \equiv \sin\theta_{n-1}+\sin\theta_{n+1} 
\nonumber \\ \approx
& -\sin\Theta (2-\vartheta^2-a^2\Theta_x^2) +\cos\Theta (2\vartheta-a^2\Theta_{xx}).
\end{align}
There are two more dipole-interaction contributions $d_2$ and $d_3$, in the 2nd and 3rd lines of
Eq.\ (\ref{dotphi}), which are expanded in Appendix \ref{App1}. The expansions of the anisotropy and field
terms are straightforward and also shown in Appendix \ref{App1}. I define a dimensionless time variable,
\be
\tau \equiv \frac{\gamma D}{\mu} t,
\ee
and use dot to indicate derivative with respect to $\tau$.
From the opposing rotations assumption, Eq.\ (\ref{map}), the discrete dynamic equations found for even-$n$ sites 
and for odd-$n$ sites are related to the desired continuum equations by the following:
\begin{align}
\label{inp}
\dot\Phi &= \tfrac{1}{2}(\dot\phi_{\text{evn}}-\dot\phi_{\text{odd}}),
\nonumber \\
\dot\varphi &= \tfrac{1}{2}(\dot\phi_{\text{evn}}+\dot\phi_{\text{odd}}).
\end{align}
$\dot\phi_{\text{evn}}$ and $\dot\phi_{\text{odd}}$ are defined by Eq.\ (\ref{dotphi}) expanded
in the continuum angles for even and odd $n$, respectively, which is done in Appendix \ref{App1}. 
From here, position $x$ will be measured in units of $a$, equivalent to setting $a=1$ in the equations. 
Then, the net continuum equation for the large in-plane angle, expanded to quadratic order 
of small terms, is,
\begin{align}
\label{dPhidt0}
\dot\Phi & \approx
\sin\Theta \cos^2\!\Phi (2-\Theta_x^2) -(1+\cos^2\!\Phi)\sin\Theta(\Phi_x^2)
\nonumber \\
& +\tan\Theta\sin\Theta \sin 2\Phi (\Theta_x\Phi_x)
 -\sin\Theta \sin 2\Phi (\tfrac{1}{2}\Phi_{xx})
\nonumber \\
& -\sec\Theta(1+\sin^2\!\Theta\cos^2\!\Phi)\Theta_{xx} -6\varphi^2 \sin\Theta
\nonumber \\
& +4\vartheta^2 \sin\Theta \left[ \sec^2\!\Theta+\left(\sec^2\!\Theta-\tfrac{1}{4}\right)\cos^2\!\Phi \right]
\nonumber \\
& +2k_1 \big[\left(1-\tfrac{1}{2}\vartheta^2\right)\sin\Theta \sin^2\!\Phi
+\varphi^2 \sin\Theta\cos 2\Phi
\nonumber \\
& +\varphi\vartheta\cos\Theta\sin 2\Phi \big]
+2k_3 \left(1-\tfrac{1}{2}\vartheta^2\right)\sin\Theta
\nonumber \\
& +b \left( \vartheta\sec^2\!\Theta\sin\Phi+\varphi \tan\Theta\cos\Phi \right) .
\end{align}
The first four lines are the dipolar terms. 
The net continuum equation for the small in-plane angle, also to quadratic order, is,
\begin{align}
\label{dphidt0}
\dot\varphi & \approx 
2\vartheta \sec\Theta\left\{2+\left(1+\sin^2\!\Theta\right)\cos^2\!\Phi\right\}
\nonumber \\
& +2k_1 \left(\vartheta\cos\Theta\sin^2\!\Phi+\varphi\sin\Theta\sin 2\Phi \right)
+2k_3 \vartheta \cos\Theta
\nonumber \\
& +b\big[ \left(1-\tfrac{1}{2}\varphi^2+\vartheta^2\sec^2\!\Theta\right)\tan\Theta\sin\Phi
\nonumber \\
& +\vartheta\varphi\sec^2\!\Theta\cos\Phi \big].
\end{align}

\subsection{Out-of-plane dynamic equations}
A similar expansion is carried out in Appendix \ref{App1} for the out-of-plane continuum large and small angles. 
The opposing rotations assumption (\ref{map}) gives the  relation connecting them to the discrete angles on
even-$n$ and odd-$n$ sites,
\begin{align}
\dot\Theta &= \tfrac{1}{2}(\dot\theta_{\text{evn}}-\dot\theta_{\text{odd}}),
\nonumber \\
\dot\vartheta &= \tfrac{1}{2}(\dot\theta_{\text{evn}}+\dot\theta_{\text{odd}}).
\end{align}
where $\dot\theta_{\text{evn}}$ and $\dot\theta_{\text{odd}}$ are defined by Eq.\ (\ref{dottheta}) expanded
in the continuum angles for even and odd $n$, respectively. The result for the large out-of-plane angle, 
to quadratic order terms, is,
\begin{align}
\label{dThetadt0}
\dot\Theta & \approx
 \cos\Theta [-\sin 2\Phi+\Phi_{xx}(2\sin^2\!\Phi+\cos^2\!\Phi)
\nonumber \\
& +\sin\Phi\cos\Phi \{\vartheta^2+\Theta_x^2+\Phi_x^2\} ]
+\sin\Theta [\Theta_{xx} \sin\Phi\cos\Phi
\nonumber \\
& -6\vartheta\varphi
 -2\Theta_x\Phi_x(2\sin^2\!\Phi+\cos^2\!\Phi)]
\nonumber \\
& +k_1 \big\{ (1-2\varphi^2-\tfrac{1}{2}\vartheta^2)\cos\Theta\sin 2\Phi
\nonumber \\
& -2\vartheta\varphi\sin\Theta\cos 2\Phi \big\} -b \varphi\sin\Phi.
\end{align}
For the small out-of-plane angle, there results,
\begin{align} 
\label{dthetadt0}
\dot\vartheta \approx &
-( 6\varphi \cos\Theta + \vartheta\sin\Theta \sin 2\Phi)
\nonumber \\
& +k_1 ( 2\varphi \cos\Theta \cos 2\Phi -\vartheta\sin\Theta \sin 2 \Phi)
\nonumber \\
& +b \left(1-\tfrac{1}{2}\varphi^2\right)\cos\Phi .
\end{align}
All four dynamic equations have considerable complexity, and up to now I have not found dynamic
DW solutions. The analysis here focusses on {\em static} solutions, for which the out-of-plane angles 
become zero.

\section{Continuum equations via a Hamiltonian density}
With the continuum dynamic equations found by expansion of the discrete ones being complicated, a
Hamiltonian approach was also used for verification.  For the NN model with range $R=1$, the
discrete Hamiltonian (\ref{Hn}) in planar spherical coordinates is
\begin{align}
\label{Ha}
H & = \sum_{n} \Big\{ D \big[ \sin\theta_n\sin\theta_{n+1} 
 +\cos\theta_n \cos\theta_{n+1} \times 
 \\
& (-2\cos\phi_n\cos\phi_{n+1}+\sin\phi_n\sin\phi_{n+1}) \big] 
\nonumber \\
& -K_1\cos^2\theta_n \sin^2\phi_n 
 +K_3\sin^2\theta_n -\mu B \cos\theta_n\sin\phi_n \Big\}.
\nonumber 
\end{align}
In the continuum theory, the Hamiltonian density ${\cal H}$ is effectively the energy per island,
defined implicitly by
\be
H = \int dx\ {\cal H}.
\ee
Using the opposing rotations assumption (\ref{map}), ${\cal H}$ can be expressed in terms of the
large and small continuum angles.  The result must be expanded to the minimum order necessary to
recover the continuum dynamic equations up to second order small terms.  For that to work out, it
is necessary to get ${\cal H}$ while including up to 3rd order small terms, because the dynamic
equations are obtained by its derivatives with respect to the angular coordinates and their gradients.

The continuum expansion of ${\cal H}$ for opposing rotations, 
found in Appendix \ref{App2}, is
\begin{align}
\label{H-or}
\frac{\cal H}{D} &= -1 +\cos^2\Theta \big[-\cos^2\Phi+(1-\tfrac{1}{2}\cos^2\!\Phi)\Phi_x^2+3\varphi^2 \big]
\nonumber \\
& +(1+\sin^2\Theta\cos^2\Phi) \tfrac{1}{2}\Theta_x^2
+\tfrac{1}{4} (\sin 2\Theta \sin 2\Phi)\, \Theta_x \Phi_x 
\nonumber \\
& +(2 +\cos^2\Phi)\vartheta^2 
\nonumber \\
&  -k_1 \big[ \cos^2\Theta(\sin^2\Phi+\varphi^2\cos 2\Phi) -\vartheta^2 \cos 2\Theta \sin^2\Phi 
\nonumber \\
& -\vartheta\varphi\sin 2\Theta\sin 2\Phi \big]
+k_3 \left(\sin^2\Theta +\vartheta^2 \cos 2\Theta \right)
\nonumber \\
& -b\big[ -\left(\vartheta-\tfrac{1}{6}\vartheta^3-\tfrac{1}{2}\vartheta\varphi^2\right)\sin\Theta\sin\Phi
\nonumber \\
& +\left(\varphi-\tfrac{1}{6}\varphi^3-\tfrac{1}{2}\varphi\vartheta^2\right)\cos\Theta\cos\Phi \big].
\end{align} 
The dipolar terms in the first three lines would be different if instead {\em opposite directions} was assumed. 
The anisotropy and field terms do not depend on the assumption.  Only the Oersted energy term
has 3rd order small terms. Symmetry eliminates them from the dipolar and anisotropy parts.

For the continuum dynamics, the Hamiltonian equations (\ref{HamEq}) need to be expressed in terms of the
angles $\Phi,\ \varphi,\ \Theta$, and $\vartheta$.  The transformation is demonstrated in Appendix \ref{App2}.
For the in-plane angles, the dynamics is
\begin{align}
\label{dot-in-or}
\dot\Phi & = 
\sec\Theta \left\{ \left[1 +\vartheta^2\left(\tan^2\!\Theta+\tfrac{1}{2}\right)\right]
\frac{\partial H}{\partial \Theta}
+\vartheta \tan\Theta \frac{\partial H}{\partial \vartheta} \right\},
\nonumber \\
\dot\varphi & = 
\sec\Theta \left\{ \vartheta\tan\Theta\frac{\partial H}{\partial \Theta}
+\left[1 +\vartheta^2\left(\tan^2\!\Theta+\tfrac{1}{2}\right)\right]
\frac{\partial H}{\partial \vartheta} \right\}.
\end{align}
The partial derivatives are interpreted as functional derivatives of the Hamiltonian, i.e.,
\be
\frac{\partial H}{\partial \Theta} \longrightarrow \frac{\delta \cal H}{\delta \Theta}
= \frac{\partial \cal H}{\partial \Theta} 
-\frac{d}{dx} \left(\frac{\partial \cal H}{\partial \Theta_x}\right).
\ee
Similar equations apply to the out-of-plane dynamics,
\begin{align}
\label{dot-out-or}
\dot\Theta & = -\sec\Theta \left\{ 
\left[1 +\vartheta^2\left(\tan^2\!\Theta+\tfrac{1}{2}\right)\right] \frac{\partial H}{\partial\Phi}
+\vartheta\tan\Theta \frac{\partial H}{\partial \varphi} \right\}.
\nonumber \\
\dot\vartheta &= -\sec\Theta 
\left\{ \vartheta\tan\Theta \frac{\partial H}{\partial\Phi}
+\left[1 +\vartheta^2\left(\tan^2\!\Theta+\tfrac{1}{2}\right)\right] \frac{\partial H}{\partial \varphi} \right\},
\end{align}
Equations of comparable form were developed for the problem of solitons in AFM chains \cite{Mik80,FlugMik83}.
When Eqs.\ (\ref{dot-in-or}) and (\ref{dot-out-or}) are applied to the Hamiltonian density (\ref{H-or}),
the dynamic equations up to second order small terms agree with those found in Sec.\ \ref{expand}, that were
obtained by expansion of the discrete equations of motion.

\section{Continuum equations for static equilibrium}
If all $\dot\phi_n=0$, the discrete dynamic equation (\ref{dotphi}) is satisfied by putting all 
$\theta_n=0$. Thus, static solutions take a planar form, which was also obtained in the numerical 
relaxation simulations.  

For statics, the continuum out-of-plane angles $\Theta,\ \vartheta$ both become zero, because $\theta_n=0$
for all even and odd sites.  This reduces the in-plane dynamic equations (\ref{dPhidt0}) and (\ref{dphidt0}) 
identically to $\dot\Phi=0$ and $\dot\varphi=0$. The out-of-plane dynamic equations (\ref{dThetadt0}) and 
(\ref{dthetadt0}) control the statics. In this planar case the {\em equilibrium equations} are
\begin{align}
\label{dThetadt1}
\dot\Theta \approx &
 -\left(1-\tfrac{1}{2}\Phi_x^2\right)\sin 2\Phi+\Phi_{xx}(2\sin^2\!\Phi+\cos^2\!\Phi)
\nonumber \\
& +k_1 (1-2\varphi^2)\sin 2\Phi -b \varphi\sin\Phi = 0.
\end{align}
\begin{align} 
\label{dthetadt1} 
\dot\vartheta \approx & -6\varphi +2k_1 \varphi\cos 2\Phi 
 +b \left(1-\tfrac{1}{2}\varphi^2\right)\cos\Phi  = 0.
\end{align}
The latter gives $\varphi$ to linear order in the field $b$,
\be
\label{vphi}
\varphi \approx \frac{b\cos\Phi}{2(3-k_1\cos 2\Phi)},
\ee
and this eliminates $\varphi$ from (\ref{dThetadt1}), which becomes the equilibrium equation
in the opposing rotations scenario,
\begin{align}
\label{EE}
\dot\Theta  = &
(k_1-1+\tfrac{1}{2}\Phi_x^2)\sin 2\Phi +(2-\cos^2\Phi) \Phi_{xx} 
\nonumber \\
& -\frac{b^2}{4}\frac{(3+k_1)\sin 2\Phi} {(3-k_1 \cos2\Phi)^2} =0.
\end{align}
This equation is the basis for finding static $y$-alt$^2$ DW solutions in the continuum theory. 
Approximate but very good solutions for the continuum in-plane structure are described next.

\section{Continuum static DWs with $b=0$}
The field $b$ breaks symmetry between the sublattices.  Initially, consider $b=0=\varphi$, and refer to the DW
in Fig.\ \ref{phi1001} found numerically. The in-plane angle $\phi_n$for the even-$n$ sublattice is $\Phi$,
which has a large slope at the DW center, but zero slope as $x\to \pm \infty$ (supposing an infinitely
long system). The almost constant slope near the center suggests a possible power-series analysis there, however, 
I did not find that fruitful.  Instead, an asymptotic analysis for $x\to \pm \infty$ works well, where Eq.\ (\ref{EE})
exhibits exponential behavior.  

For the asymptotic analytics, I consider a DW centered at the origin, with an approximate half-width $h$. 
The angle $\Phi$ can take asymptotic values $\Phi(-\infty) = \pm\frac{\pi}{2}$, $\Phi(+\infty) = \mp\frac{\pi}{2}$,
with opposite signs at the two limits. To be specific to the case in Fig.\ \ref{phi1001}, introduce a function 
$\psi(x) \ge 0$ with $\psi(x)\ll 1$ at $|x|\gg h$, such that
\be
\label{Phix}
\Phi(x) = \begin{cases} +\frac{\pi}{2} -\psi(x), & x < 0, \\ -\frac{\pi}{2}+\psi(x), & x \ge  0. \end{cases}
\ee
For negative and positive $x$, this produces
\begin{align} 
\sin 2\Phi & = -\sin 2\psi\, \sgn(x) \approx -2\psi\, \sgn(x),
\nonumber \\
\cos\Phi & = \sin\psi \approx \psi,
\quad
\Phi_{xx} = \sgn(x) \psi_{xx}.
\end{align}
The opposing rotations equilibrium equation (\ref{EE}) becomes approximately,
\be
\label{asymp}
(2-\psi^2) \psi_{xx} -(k_1-1+\tfrac{1}{2}\psi_x^2)(2\psi) \approx 0.
\ee
Both factors depended on the signum function, which then divided out.  For $|x| > h$, I consider $\psi$ itself 
to be 1st order small; then the terms $\psi \psi_x^2$ and $\psi^2\psi_{xx}$ are of 5th order and can be 
neglected in the asymptotic solution.  This results in the {\em linearized} equilibrium equation, 
extended to all $x$,
\be
\label{linear}
\psi_{xx} \approx (k_1-1)\psi,
\ee
where these terms are of 3rd order smallness. On each side of the origin, this has an exponentially
decaying solution that represents the DW.  The other exponentially growing solutions are unstable and therefore 
unphysical. At the origin, I take $\Phi(0)=0$, which gives an initial value $\psi(0)=\frac{\pi}{2}$ for the DW 
in Fig.\ \ref{phi1001}. Then the solution of the linearized equation for $\psi$ is
\be
\label{psix}
\psi(x) = \frac{\pi}{2} e^{-\beta |x|}, \quad \beta \equiv \sqrt{k_1-1} = h^{-1}.
\ee
The parameter $\beta$ can be identified as an inverse half-width $h$ for the DW.
Although I assumed asymptotically large $|x|$ to derive this, $\psi$ has exactly the correct value at the DW center,
and its physically useful range covers the entire $x$-axis. This implies for the large in-plane angle,
\be
\label{DW}
\Phi(x) = -\sgn(x) \frac{\pi}{2} \left(1-e^{-\beta |x|} \right).
\ee
This describes the DW in Fig.\ \ref{phi1001}. A DW can be constructed in a total of
four different symmetries. A second solution is obtained by reversing the sign of $\Phi$ in (\ref{DW}).
There are two more solutions where the dipoles point at $\Phi(0)=\pm\pi$ at the DW center, described by
\be
\label{DW1}
\Phi(x) = \pm \frac{\pi}{2} \left(1+e^{-\beta |x|} \right).
\ee
The four possible DW symmetries represent the four ways the dipoles can rotate from one transverse direction
(i.e., $S^y=\pm 1$) to the opposite transverse direction ($S^y=\mp 1$) while the sublattices follow the 
opposing rotations scenario.

\begin{figure}
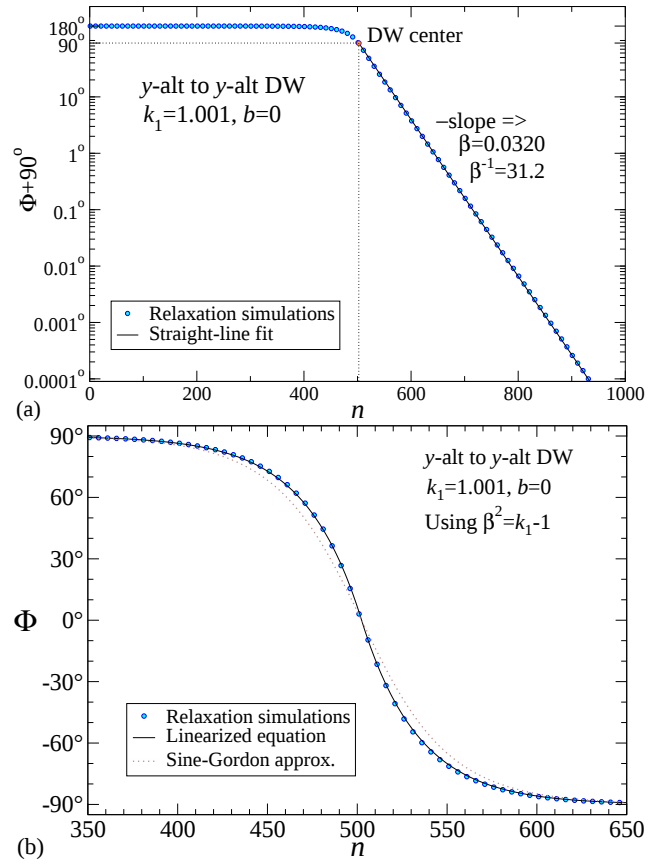

\includegraphics[width=\figwidth,angle=0]{logphi-k1001}
\includegraphics[width=\figwidth,angle=0]{yalt+falt-k1001}
\caption{\label{DW-k1001} (a) For $k_1=1.001$, $N=1000$, alternating order parameter $\phi_{\rm alt}$ for a
$y$-alt$^2$ DW from the numerical simulations (blue circles) is used to estimate inverse half-width $\beta$ from
the negative slope of the log-linear plot,  based on Eq.\ (\ref{lleqn}). Only every tenth site is plotted.
The asymptotic function is the black straight line drawn only
to the right side of the DW center. The slope came from the DW center
point, $\phi_{\rm alt}(502.4)=0$, and the far-field point, $\phi_{\rm alt}(934)+90=9.0\times 10^{-5}$.
Theory Eq.\ (\ref{psix}) gives $\beta=\sqrt{0.001} \approx 0.03162$.
(b) Zoomed-in alternating order $\phi_{\rm alt}$ of the DW on a linear plot.
The numerical simulations are compared with the linear equilibrium Eq.\
(\ref{linear}) and the sine-Gordon approximation in Eq.\ (\ref{sG}).
Only every fifth site of the numerical data is plotted.}
\end{figure}

For $b=0$, the angle $\Phi$ is the same as the discrete in-plane angle $\phi_{\rm evn}$ on the even sites;
the odd sites take the opposite sign. In order to compare to the numerical solutions, it is good to define
an alternating order parameter, $\phi_{\rm alt}$, that extracts the effective value of $\Phi$ from the discrete 
data. In a continuum definition, this would be
\be
\phi_{\rm alt}(x) \equiv \tfrac{1}{2} \left[\phi_{\rm evn}(x)-\phi_{\rm odd}(x)\right].
\ee 
To implement this symmetrically on a grid of sites, the average neighbor sites were subtracted from
a central site, as in
\be
\label{phialt}
\phi_{{\rm alt}, n} = \tfrac{1}{2}[\phi_n-\tfrac{1}{2}(\phi_{n-1}+\phi_{n+1})]
\times
\begin{cases} +1, & n=\text{even}, \\ -1, & n=\text{odd}. \end{cases}
\ee
A nonsymmetric difference was used at the chain ends.  $\phi_{\rm alt}$ is the discrete estimate of $\Phi$
for the configuration.  If there is no alternating order, $\phi_{\rm alt}$ equals zero.

To see how well the continuum solution in Eq.\ (\ref{DW}) represents the numerical solutions, it is fruitful to
zoom in on the asymptotic behavior for $x\to +\infty$ via a logarithmic transformation. Eq.\ (\ref{DW}) is
equivalent to expecting the large in-plane angle $\Phi$ (or the numerically calculated $\phi_{\rm alt}$) 
to behave as
\be
\label{lleqn}
\ln\left(\frac{\Phi+\frac{\pi}{2}}{\frac{\pi}{2}}\right) = -\beta x, \quad x>0.
\ee
The logarithmic function on the left side also should pass through zero at the DW center, and tend towards 
$\ln 2$ as $x \to -\infty$. The function can equally be defined using $\Phi$ in degrees, changing 
$\frac{\pi}{2}$ to $90^{\circ}$. 

For a DW with $k_1=1.001$ and $b=0$, the asymptotic behavior is displayed in Fig.\ \ref{DW-k1001}a
in a log-linear plot. The fitted straight line decay slope ($\beta=0.0320$) is surprisingly close to the
continuum prediction from Eq.\ (\ref{psix}), $\beta=0.03162$. The inverse of $\beta$ defines the DW
fitted half-width $h\approx 31.2$ (in lattice constants) and the continuum prediction is 
$h \approx \sqrt{1/0.001} = 31.6$, very nearly the same. This large width helps to minimize discreteness
effects and make the continuum analysis reliable.

Of course, the log-linear plot disguises the behavior near the DW center. The linearized theory curve for $\Phi$ 
from Eq.\ (\ref{DW}) is shown in Fig.\ \ref{DW-k1001}b, compared to the numerical data and a sine-Gordon
approximation, Eq.\ (\ref{sG}), described below.  The linear theory curve fits well at the DW center,
and also for $|x| \to \pm\infty$. The regions with the largest deviation of the theory from the numerical 
calculation are in the shoulders of the DW, about a distance $h$ from the center, but even those deviations
are extremely small.   

\begin{figure}
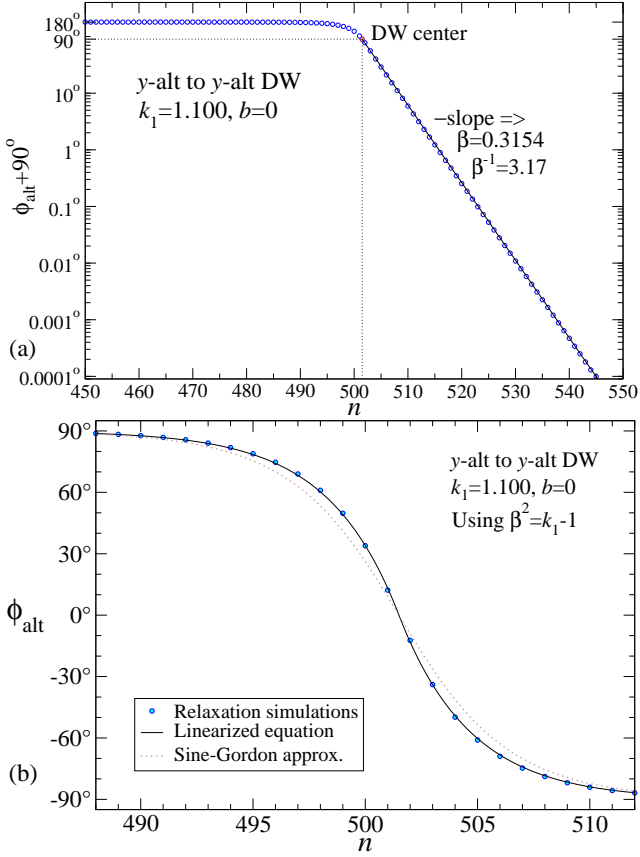

\includegraphics[width=\figwidth,angle=0]{logphi-k11}
\includegraphics[width=\figwidth,angle=0]{yalt+falt-k11}
\caption{\label{DW-k11} (a) For $k_1=1.100$, $b=0$, $N=1000$, alternating order $\phi_{\rm alt}(x)$ in a 
$y$-alt$^2$ DW from the numerical simulations (blue circles) is used to estimate $\beta$ from the negative slope 
of the log-linear plot, based on Eq.\ (\ref{lleqn}). The asymptotic function is the black straight line drawn only
to the right side of the DW center. The slope came from the DW center
point, $\phi_{\rm alt}(501.5)=0$, and the far-field point, $\phi_{\rm alt}(545.3)+90=9.0\times 10^{-5}$.
Theory Eq.\ (\ref{psix}) gives $\beta=\sqrt{0.1} \approx 0.3162$.
(b) Zoomed-in view on a linear plot comparing the numerical solution with the linear equilibrium Eq.\ 
(\ref{linear}) and the sine-Gordon approximation in Eq.\ (\ref{sG}).}
\end{figure} 

Another static DW was calculated for $N=1000$, $b=0$, and anisotropy $k_1=1.100$, shown in Fig.\ \ref{DW-k11}.
Although $k_1$ is barely changed from the first example, the factor $k_1-1$ has changed by a factor of 100,
which is expected to reduce the half-width by a factor of 1/10$^{\rm th}$. The asymptotic numerical fit produces 
$h=\beta^{-1} \approx 3.17$ while both the linear and sine-Gordon theories predict 
$h=1/\sqrt{0.100}\approx 3.162$, although it is clear that the sine-Gordon approximation fails in the shoulders
of the DW.  It is possible to consider larger values of $k_1$ up to $k_1\approx 1.75$, however, by then the
DW acquires a fixed minimum half width $h_{\rm min} \approx 1.5$ and a fixed shape and energy, independent of $k_1$ 
(see Sec.\ \ref{create0}).

\subsection{A sine-Gordon approximation?}
A comparison was made to a sine-Gordon form for the DWs, which actually deviates significantly from the numerical 
solutions.  It is obtained by keeping $\sin 2\psi$ intact instead of linearizing it.  Eq.\ (\ref{linear}) is 
replaced by 
\be 
\label{sG} 
(k_1-1)\sin(2\psi) \approx  2\psi_{xx} .
\ee
One might expect this gives more accuracy for determination of the DW. This is a static sine-Gordon equation 
\cite{sGreview} for the variable $2\psi$. It has static kink solutions.  One form that decays away for large 
positive $x$ is
\be
2\psi = 4\tan^{-1}(e^{-\beta x}), \quad \beta=\sqrt{k_1-1}.
\ee
In this formula, $\psi$ ranges from $\pi$ to 0 as $x$ goes from $-\infty$ to $+\infty$.
The net result for $\Phi$ is
\begin{align}
\label{sGkink}
\Phi_{\rm sG} = -\frac{\pi}{2}+\psi & =  -\frac{\pi}{2} +2\tan^{-1}(e^{-\beta x})
\nonumber \\
& = -2 \tan^{-1}\left( \tanh \tfrac{1}{2} \beta x \right).
\end{align}
This applies for $-\infty<x< +\infty$, without absolute value on $x$.
The last expression was found using the identity for a tangent of a difference of angles.  While being 
interesting, it does not closely follow the DW shapes for this model, especially at the shoulders, 
as seen in Figs.\ \ref{DW-k1001} and \ref{DW-k11}. Full linearization must provide a fortuitous cancellation 
of errors relative to the nonlinear equation (\ref{EE}).

\subsection{DW creation energy for $b=0$}
\label{create0}
The DW creation energy $E^*$ is the energy needed to push a DW into the system initially in a uniform $y$-alt state.
I take it as the difference of the system energy with a DW minus the energy of a uniform $y$-alt state.
For a static DW, the energy density from Eq.\ (\ref{H-or}) with $\Theta=\vartheta=0$ and $b=0$ is
\be
\frac{\cal H}{D} = -1-\cos^2\Phi+({1-\tfrac{1}{2}\cos^2\Phi})\Phi_x^2 -k_1 (1-\cos^2\!\Phi).
\ee
This is expressed in terms of $\cos\Phi=\sin\psi$ [see Eq.\ (\ref{Phix})] for ease of calculation, which gives
\begin{align}
\frac{\cal H}{D} &= -(1+k_1)+(k_1-1)\sin^2\!\psi+(1-\tfrac{1}{2}\sin^2\!\psi)\psi_x^2.
\end{align}
In the solution (\ref{DW}), $\psi$ ranges from 0 to $\pi/2$, ruling out expansion of ${\cal H}$ 
for small or large $|\psi|$.  Instead, I appeal to numerical evaluations of the needed integrals. When $\psi=0$, 
the first term in the expression is the energy density of a uniform $y$-alt state,
\be
{\cal H}_{y-\text{alt}} / D = -(1+k_1).
\ee 
The remaining parts of ${\cal H}$ give the creation energy,
\be
\frac{E^*}{D} = \int_{-\infty}^{\infty} dx 
\left\{ (k_1-1)\sin^2\!\psi+(1-\tfrac{1}{2}\sin^2\!\psi)\psi_x^2 \right\}.
\ee
With $\psi$ as in Eq.\ (\ref{psix}) and $\psi_x^2 = \beta^2 \psi^2$, the needed integrals are:
\begin{align}
\label{int123}
\int_{-\infty}^{\infty} dx\ \psi^2 & 
= \left(\frac{\pi}{2}\right)^2\beta^{-1}.
\nonumber \\
\int_{-\infty}^{\infty} dx\ \sin^2\!\psi & 
\approx 1.6482776\ \beta^{-1}.
\nonumber \\
\int_{-\infty}^{\infty} dx\ \psi^2\, \sin^2\!\psi &
\approx 1.733700\, \beta^{-1}.
\end{align}
Then the net creation energy is predicted to be 
\be
\label{EDW}
E^*/D \approx 3.2488287\, \beta \approx 3.2488 \sqrt{k_1-1} .
\ee
Note especially that this does not depend on the length $N$ of the system. The contribution to the
energy density is concentrated mostly from shoulder to shoulder in the DW.  The formula should work
well in a discrete system much longer than the DW width.  As $\beta=h^{-1}$, the creation energy 
is inversely proportional to the DW width. The formula also shows that $k_1=1$ is a critical point
below which there are no $y$-alt$^2$ DWs.

\begin{figure}
\includegraphics[width=\figwidth,angle=0]{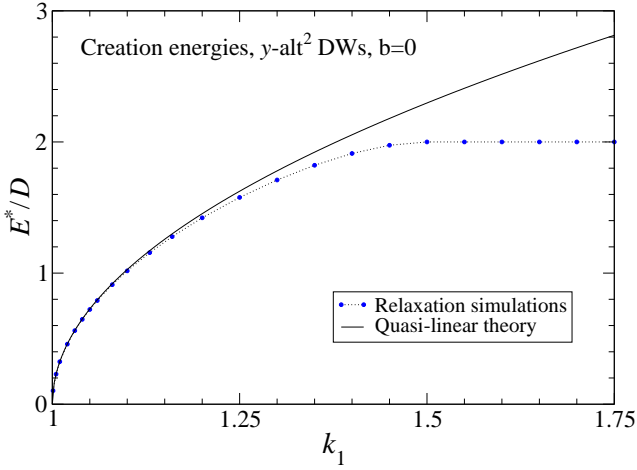}
\caption{\label{figEDW} The $y$-alt$^2$ DW creation energy vs.\ anisotropy $k_1$ from simulations (circles)
compared with the quasi-linear theory expression in Eq.\ (\ref{EDW}) (solid black curve), at zero applied
field. For $k_1\ge 1.5$, the DW acquires a minimum width $w=3a$ and a fixed energy $E^{*} = 2 D$,
as well as a shape independent of $k_1$. For $k_1 < 1.25$ the continuum theory works extremely well because 
the DW width $w=2/\sqrt{k_1-1}$ contains many sites.}
\end{figure}

For the first example in Fig.\ \ref{phi1001} with $k_1=1.001$, the formula gives 
$E^{*} \approx 0.10274 \, D$, while the numerical simulations produced $E^{*} \approx 0.10265 \, D$. The value 
from numerics was found using a system with $N=1000$, and subtracting the total energies with and without
the DW present in a surrounding $y$-alt background.  It is remarkable that although $\psi$ was obtained 
from a linearized equation, the energy must be calculated using that linearized but
very good solution, taking into account the nonlinearity in the sine functions within ${\cal H}$.
I refer to this and expression (\ref{EDW}) as the quasi-linear theory for the DW energy.

Further creation energies for a range of $k_1>1$ are exhibited in Fig.\ \ref{figEDW}. For $k_1 < 1.25$, 
the linearized continuum theory for $E^*$ is only a few percent higher than the values from simulations.
In that region, the DW width $w = 2h \ge 4$, so that the structure is reliably described using the 
continuum approach. In contrast, for $k_1\ge 1.5$, the creation energy found from simulations becomes a 
constant $E^{*}= 2D$, and the DW takes on a particular discrete shape, of a minimum full-width  $w=3$, 
which cannot be represented adequately in the continuum theory.

\section{Continuum static DWs with nonzero magnetic field}
A transverse applied magnetic field $b=\mu B/D$ tends to impart an aligned or ferromagnetic-like order to the DW,
which should obey the equilibrium Eq.\ (\ref{EE}).  This can be analyzed starting again with the large-$|x|$ 
transformation to variable $\psi(x) \ge 0$ in Eq.\ (\ref{Phix}). Assuming $\psi^2 \ll 1$, one has
$\cos 2\Phi \approx 3+k_1$ and the field contribution to the equilibrium equation is simplified, giving
\be
(2-\psi^2)\psi_{xx} -\left[k_1-1-\frac{b^2}{4(3+k_1)}+\tfrac{1}{2}\psi_x^2\right](2\psi) \approx 0.
\ee
Asymptotically, the fifth order small terms can be dropped, and the DW is still described by a linearized 
equilibrium equation, 
\be
\psi_{xx} \approx  \left[k_1-1-\frac{b^2}{4(3+k_1)}\right]\psi .
\ee
The field term effectively subtracts from the $k_1$ anisotropy.  The solution is similar to that
for zero field, but with a modified inverse half-width parameter,
\be
\label{psi-b}
\psi = \frac{\pi}{2} e^{-\beta |x|}, \quad \beta = \sqrt{k_1-1-\frac{b^2}{4(3+k_1)}}=h^{-1}.
\ee

\begin{figure}
\includegraphics[width=\figwidth,angle=0]{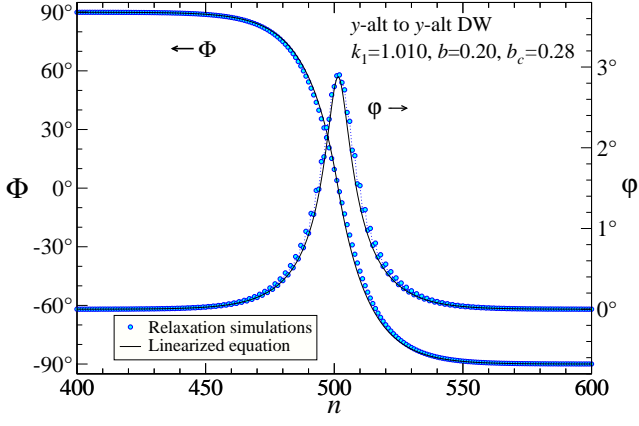}
\caption{\label{kink101b02} For $k_1=1.01$ and $b=0.20$, a $y$-alt to $y$-alt DW 
from the numerical simulations with $N=1000$, is compared with the theoretical
shape from Eqs.\ (\ref{DW}) and (\ref{vphi+b}).
For the theory curves, the DW center was placed at $x=502$ to match the data.} 
\end{figure}

A typical DW is described by Eq.\ (\ref{DW}) or one of the other symmetries. $\varphi$ is then
determined by Eq.\ (\ref{vphi}) or equivalently,
\be
\label{vphi+b}
\varphi = \frac{b\sin\psi}{2(3+k_1 \cos 2\psi)}.
\ee
At the DW center, where $\Phi=0$ and $\psi=\frac{\pi}{2}$, the maximum tilt towards the applied field is
\be
\varphi(0) = \varphi_{\rm max} \approx \frac{b}{2(3-k_1)}.
\ee
To insure real $\beta$, the field must be less than an upper critical limit, 
\be
b_c = \sqrt{4(k_1-1)(3-k_1)}.
\ee 
For example, when $k_1=1.25$, one has $b_c=1.32$, and then $\varphi_{\rm max}\approx 0.38$ rad. That is at
a stability limit and a more typical value for a stable $y$-alt$^2$ DW is only a few degrees.  

\begin{figure}
\includegraphics[width=\figwidth,angle=0]{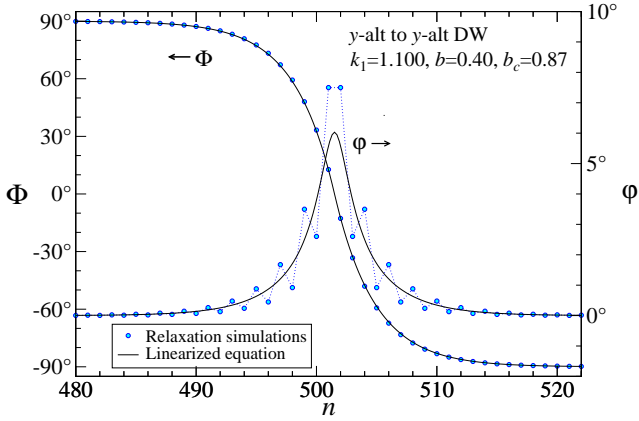}
\caption{\label{kink11b04} For $k_1=1.100$ and $b=0.40$, a $y$-alt to $y$-alt DW 
from the numerical simulations with $N=1000$, is compared with the theoretical
shape from Eqs.\ (\ref{DW}) and (\ref{vphi+b}).
A dotted line is used to distinguish the simulation data for $\varphi$.
For the theory curves, the DW center was placed at $x=502$ to match the data.} 
\end{figure}

To compare with numerical simulations, both the alternating order and aligned order is to be extracted
from a relaxed DW configuration. The ferromagnetic-like order can be estimated at a site by a slight
modification of the formula (\ref{phialt}) used to find the alternating order. We use an average
over the sublattices, 
\be
\label{phifm}
\phi_{{\rm fm},n} = \tfrac{1}{2}[\phi_n+\tfrac{1}{2}(\phi_{n-1}+\phi_{n+1})].
\ee
This is the discrete measurement on the simulation data, to be directly compared to the
analytic estimate of $\varphi$. 

The theory matches the simulations exceptionally well at $k_1=1.001$ as used in an earlier example with $b=0$.
To see some slights deviations between theory and simulations, an example of a DW for $k_1=1.010$, $b_c=0.2821$, 
and $b=0.20$ is shown in Fig.\ \ref{kink101b02}.  The agreement is still very good, although the central peak in 
the theory curve for $\varphi$  is slightly narrower than the central peak in the simulation data. A site-by-site
variation in $\varphi$ is also noticeable as there are slight discreteness effects in the data.

It is interesting to check another example with $k_1=1.100$, $b_c=0.8718$, and $b=0.40$ as shown in 
Fig.\ \ref{kink11b04}. The larger field creates a greater peak deviation.  Discreteness effects are strong 
and the smooth continuum theory for $\varphi$ passes through the average of the site-to-site oscillations of 
the simulation data, except at the peak.  As long as $b$ is not too close to the critical limit $b_c$, these 
examples show that the continuum theory using the linearized equilibrium equation is good at describing the 
DW structures.

\subsection{Longitudinal magnetic moment $m_x$ of a DW}
A $y$-alt$^2$ DW has a net magnetic moment $m_x$ parallel to the chain direction, with positive and
negative values that are only slightly dependent upon the applied field $b$.  The discrete and 
continuum definitions in spin units are
\be
m_x \equiv \sum_n S_n^x = \int_{-\infty}^{+\infty} \frac{dx}{a}\ \langle S_n^x \rangle  .
\ee 
The average is over the sublattices, both of which have nearly the same $S_n^x$.  In the continuum 
opposing rotations scenario for the two sublattices, they are
\be
S_n^x = \cos(\pm \Phi +\varphi) \approx \cos\Phi\left(1-\tfrac{1}{2}\varphi^2\right)-\sin(\pm\Phi)(\varphi).
\ee
Then the average is
\be
\langle S_n^x \rangle \approx \cos\Phi\left(1-\tfrac{1}{2}\varphi^2\right).
\ee
In the DW, the asymptotic solution has $\cos\Phi = \sin\psi$, and
using the value already found for $\varphi$ in Eq.\ (\ref{vphi+b}), we have
\be
\langle S_n^x \rangle \approx \sin\psi\left[1-\frac{b^2\sin^2\!\psi}{8(3+k_1\cos 2\psi)^2}\right].
\ee
Using $\cos 2\psi \approx 1$, the net $m_x$ depends on integrals over $x$ or over $\psi$ that are 
evaluated numerically:
\begin{align}
\int_{-\infty}^{\infty} dx\, \sin\psi & = 2\int_0^{\frac{\pi}{2}} \frac{d\psi}{|d\psi/dx|} \sin\psi
\approx  2.741 \beta^{-1}.
\nonumber \\
\int_{-\infty}^{\infty} dx\, \sin^3\!\psi & =  2\int_0^{\frac{\pi}{2}} \frac{d\psi}{|d\psi/dx|} \sin^3\!\psi
\approx  1.252 \beta^{-1}.
\end{align}
That gives the estimate for the magnitude,
\be
\label{mxa}
|m_x| \approx \left[2.74-\frac{1.25 b^2}{8(3+k_1)^2}\right]\beta^{-1} \approx 2.74 \beta^{-1}.
\ee
With $k_1\approx 1$, the term with $b^2$ is negligible, although $\beta$ has implicit $b$-dependence.
A positive or negative value of $m_x$ is allowed, slightly more than the DW width. This is fairly large and 
could be used to distinguish the DW-state from a uniform $y$-alt state with $m_x=0$.   

\subsection{Transverse magnetic moment of a DW}
The magnetic field induces a net transverse magnetic moment in the dipoles in a DW.
The net induced moment is defined in discrete and continuum views by
\be
m_y \equiv \sum_n S_n^y = \int_{-\infty}^{+\infty} dx\ \langle S_n^y \rangle  .
\ee
Outside the DW, $S^y$ decays exponentially. With the main contribution from the center of the DW, 
the asymptotic DW solution (\ref{psi-b}) can give an estimate.  From the calculation of the continuum 
Hamiltonian and Eq.\ (\ref{SyAB}) with $\Theta=\vartheta=0$, the 
transverse spin component averaged over the sublattices is
\be
\langle S_n^y \rangle \approx \varphi \cos\Phi = \frac{b\sin^2\!\psi}{2(3+k_1 \cos 2\psi)}.
\ee
Integration over $x$ can be changed to integration over $\psi$, together with using the asymptotic
approximations $\sin\psi\approx \psi$ and $\cos 2\psi \approx 1$. This gives the net linearized 
continuum prediction,
\be
\label{myl}
m_y \approx 2 \int_{\psi=0}^{\psi=\pi/2} \frac{d\psi}{|d\psi/dx|} \frac{b\psi^2}{2(3+k_1)}
= \frac{\pi^2 b}{8(3+k_1)\beta}.
\ee 
Note that $\psi^2$ is an overestimate of $\sin^2\!\psi$ in the numerator, while the denominator
is also overestimated; these errors tend to cancel.

\begin{figure}
\includegraphics[width=\figwidth,angle=0]{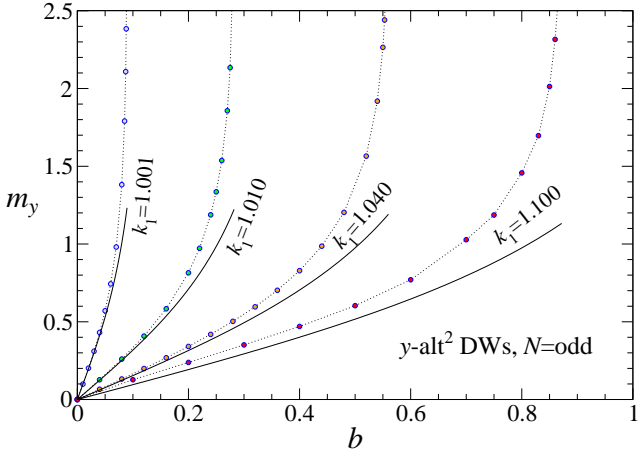}
\caption{\label{myodd} The induced transverse magnetic moment $m_y$ of a $y$-alt$^2$ DW in applied
magnetic field $b$, for the indicated anisotropy strengths.  Points are from simulations of odd-numbered
chains ($N=501$), dotted lines are guides to the eye, and the solid black curves are the linearized
theory expression in Eq.\ (\ref{myl}). For an even-$N$ chain, the results for $m_y$ are shifted
from these by the quantized intrinsic moment, $m_y=\pm 1$.}
\end{figure}

For the example with $k_1=1.01, b=0.20$, one gets $\beta = 0.0866$, and the formula gives $m_y \approx 0.710$,
while the simulated values are interesting. If the system contains an odd number of sites (say, $N=501$), the
simulated value is $m_y = 0.815$, a component aligned paramagnetically with the field, but somewhat larger
than the theory prediction. A comparison of $m_y$ obtained from the simulations and the theory for odd-$N$ is 
shown in Fig.\ \ref{myodd}. Generally the theory expression underestimates $m_y$, and works best for the
smaller values of $k_1$.  

If instead the system contains an even number of sites (say, $N=500$), the simulated value 
can be either $m_y = 1.815 = 1+0.815$ or $m_y = -0.185 = -1+0.815$, depending on the initial state of the
simulation. In both cases, the DW remains in the center of the system.  The negative value of $m_y$ means 
it points opposite to the applied field. The two cases correspond to opposite directions of rotation in
the spins as one moves down the chain.  The magnetic moment is shifted from that for the odd-$N$ system
by a quantized topological value, $m_y^{\rm topo}=\pm 1$.

Surprisingly, if $b=0$, the DW in an even-$N$ chain can have one of the exactly quantized intrinsic values, 
$m_y = m_y^{\rm topo} = \pm 1$.  This topological effect is not correctly accounted for in the continuum theory. 
It does not appear in odd-$N$ chains.  Eq.\ (\ref{myl}) for $m_y$ induced by $b$ applies directly only for 
odd-$N$ chains.  For even-$N$ chains,  Eq.\ (\ref{myl}) represents the field-induced magnetic moment, 
that takes place on top of the intrinsic values of $m_y^{\rm topo}=\pm 1$.  To get $m_y$ for even-$N$ chains, 
one can take the values plotted in Fig.\ \ref{myodd} and shift them by the intrinsic values, $\pm 1$. 

\subsection{On odd-length vs.\ even-length chains}
The odd/even chain effect is accounted for by carefully examining the configurations in
short-chain DWs ($N=49$ or $50$).

\begin{figure}
\includegraphics[width=\figwidth,angle=0]{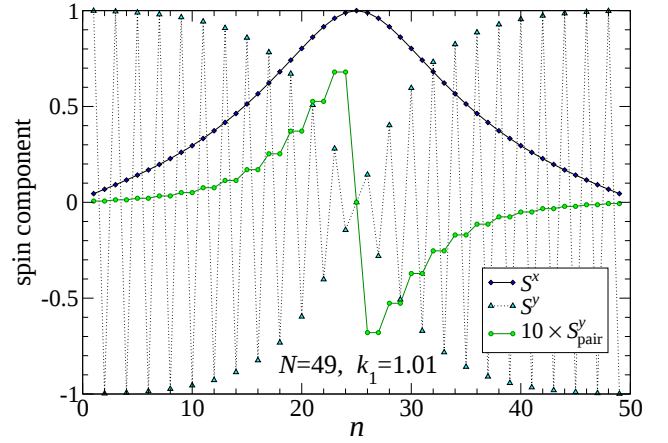}
\caption{\label{S49} A DW spin configuration for odd system length $N=49$, anisotropy $k_1=1.01$,
field $b=0$, from relaxation simulations.  $S_n^x$ and $S_n^y$ are the raw components on the islands.
Note the alternating structure present in $S_n^y$. The values of $S_{\text{pair},n}^y$ are the averages of
odd/even neighboring sites in each two-site pair, but the DW center site with $S_n^y=0$ at $n=25$
is plotted individually. The antisymmetry of $S_{\text{pair},n}^y$ leads to net $m_y=0$.}
\end{figure}

In an odd chain with $b=0$, there is a centered site with $S_n^y=0$, around which $S_n^y$
holds antisymmetry on each sublattice. An example is shown in Fig.\ \ref{S49} for $N=49$, $k_1=1.01$,
where the raw spin components $S_n^x, S_n^y$ on the islands are shown. $S_n^x$ is the same for
both sublattices, but $S_n^y$ alternates from site to site. It is difficult to visualize the
sum of $S_n^y$ to get the net $m_y$. I calculated a pair average to make it apparent. 
The average is taken over pairs of neighboring odd/even sites,
\be
S_{\text{pair},n}^y = S_{\text{pair},n+1}^y = \tfrac{1}{2}(S_n^y+S_{n+1}).
\ee
Each site in a pair is given the same average value. With $N=$odd, one site is not in a pair; it
is taken as the site at the center of the DW, which happens also to have $S_{25}^y=0$.  On the left side
of the center, the pairs are odd-even, or ($n$)-($n+1$) being 1-2, 3-4, 5-6, etc.  On the right side of the
center, the pairs are even-odd, or ($n$)-($n+1$) being 26-27, 28-29, 30-31, etc. Now the sum is
clear, and the two antisymmetric halves can be seen to cancel out and give
$m_y=\sum S_{\text{pair},n}^y=0$. 

%
\begin{figure}
\includegraphics[width=\figwidth,angle=0]{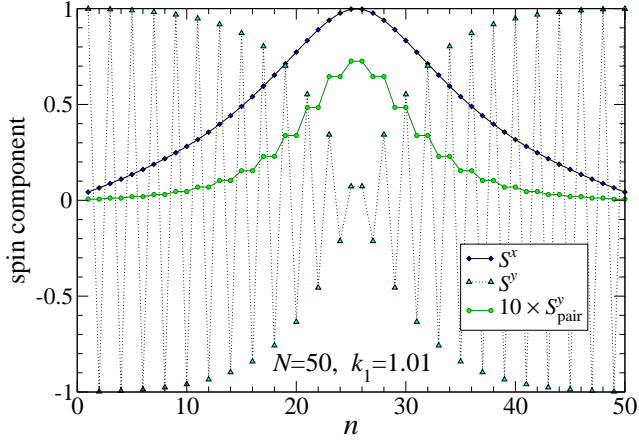}
\caption{\label{S50} A DW spin configuration for even system length $N=50$, anisotropy $k_1=1.01$,
field $b=0$, from relaxation simulations. See explanation of curves in  Fig.\ \ref{S49} for $N=49$.
Now all sites are in pairs, and $S_{\text{pair}}^y$ is symmetric around the DW center.
The net intrinsic magnetic moment is $m_y=1$, even though $b=0$.}
\end{figure}

In an even chain with $b=0$, the center of the DW falls between two sites.  An example with
$N=50$, $k_1=1.01$, and $b=0$ is shown in Fig.\ \ref{S50}, displaying the raw $S_n^x$, $S_n^y$ and
the pair-averaged $S_{\text{pair},n}^y$.  In this case, all of the pairs are of the form odd-even, or
1-2, 3-4, 5-6, etc., across the whole length of the system.  There is no site not in a pair.
It is clear that $S_{\text{pair},n}^y$ now has right/left symmetry around the DW center,
and the two halves make equal contributions to the net $m_y$ even though $b=0$. 
It creates a nonzero, quantized, intrinsic magnetic moment $m_y = \pm 1$.
This is the topological effect, caused by the addition of one (up or down) spin to an odd-$N$ chain. 
If an odd-$N$ chain has $m_y=0$, then the even-$N$ chain must have $m_y=m_y^{\rm topo}=\pm 1$, because 
the spins are all $S_n^y=\pm 1$ in the $y$-alt states at the chain ends, far from the DW center.


This way of summing by pairs, and leaving the central site of an odd-$N$ system unpaired, is a way to 
move all the contributions to $m_y$ away from the boundaries, which are in $y$-alt states. It makes the 
odd/even influence come from the DW itself.  In this sense, a boundary effect has been recast as a 
surprising geometric property of the DW.

This method contradicts the original approach of the continuum limit, because what appears in these 
pairs depends on where their construction starts. A shift by 1 site leaves some spin unpaired.  
%
%
Further, the intrinsic magnetic moment for even-$N$ systems does not emerge in the original continuum limit, 
even though it seems to come mostly from the DW central region. Instead, $S_{\text{pair},n}^y$ can
be averaged over sublattices and summed to produce the intrinsic part of $m_y$. 

The continuum version of $S_{\text{pair},n}^y = S_{\text{pair},n+1}^y$ depends
on whether $n$ is even or odd. First, consider $n=$even, and use the opposing rotations assumption
in the planar case,
\begin{align}
S_{\text{pair},n}^y & = S_{\text{pair},n+1}^y = \tfrac{1}{2}(S_n^y+S_{n+1}^y) 
\nonumber \\
& \to \tfrac{1}{2} \left[ \sin\Phi(x) +\sin(-\Phi(x+a))\right]
\nonumber \\
& = \tfrac{1}{2} \left[ \sin\Phi -\sin(\Phi +a\Phi_x +\tfrac{1}{2}a^2\Phi_{xx}) \right]
\nonumber \\
& \approx -\tfrac{1}{2} a \Phi_x \cos\Phi, \quad n=\text{even}.
\end{align}
If $n=$odd, the sign of $\Phi$ is reversed, leaving
\be
S_{\text{pair},n}^y = S_{\text{pair},n+1}^y
\approx \tfrac{1}{2} a \Phi_x \cos\Phi, \quad n=\text{odd}.
\ee
Thus, an odd-even pair has a different sign from an even-odd pair. This explains the antisymmetry
seen in the odd-$N$ chain, which has odd-even pairs on one side of the center and even-odd pairs
on the other side.  For the even-$N$ chain, all the pairs are odd-even (use $n=$odd formula) and then 
right-left symmetry is the result.

Thus, we can use the result and the asymptotic DW solution to find the intrinsic magnetic moment
for even-$N$ chains.  Consider a DW with $\Phi(-\infty)=-\frac{\pi}{2}$ and $\Phi(+\infty)=\frac{\pi}{2}$.
The continuum expression for $S_{\text{pair},n}^y$ with $n=$odd is a perfect differential, so its integration
does not depend on the details of the DW shape.  This is the reason why the intrinsic $m_y$ is
quantized, and the value is obtained by
\begin{align}
m_y^{\rm topo}  & = \sum_n S_{\text{pair},n}^y
\to \int_{-\infty}^{\infty} \frac{dx}{a}\, \tfrac{1}{2} a \frac{d\Phi}{dx} \cos\Phi
\nonumber \\
& = \tfrac{1}{2} \int_{-\infty}^{\infty} dx\, \frac{d}{dx} (\sin\Phi)
\nonumber \\
& = \tfrac{1}{2} \left[\sin\Phi(\infty)-\sin\Phi(-\infty)\right]
= 1.
\end{align}
With the DW flipped (reverse all spins), one instead gets $m_y^{\rm topo}=-1$.
This quantization of $m_y$ for even-$N$ chains with $b=0$ is indeed exact.

If it is an odd-$N$ chain, the left half of the system uses $S_{\text{pair},n}^y$ with $n=$odd and
the right half uses it with $n=$even.  They have reversed signs, and their left vs.\ right side sums cancel
out upon integration over the whole chain, as seen in Fig.\ \ref{S49}. The odd-$N$ chains have intrinsic 
$m_y=0$, exactly.

When $b\ne 0$, the continuum prediction of $m_y$, Eq.\ (\ref{myl}), is superimposed onto the intrinsic value,  
whether it be an odd-$N$ or even-$N$ chain.  This analysis explains how topological effects influence
the values of $m_y$ differently for even-$N$ vs.\ odd-$N$ chains.

\subsection{DW creation energy for $b\ne 0$}

%
\begin{figure}
\includegraphics[width=\figwidth,angle=0]{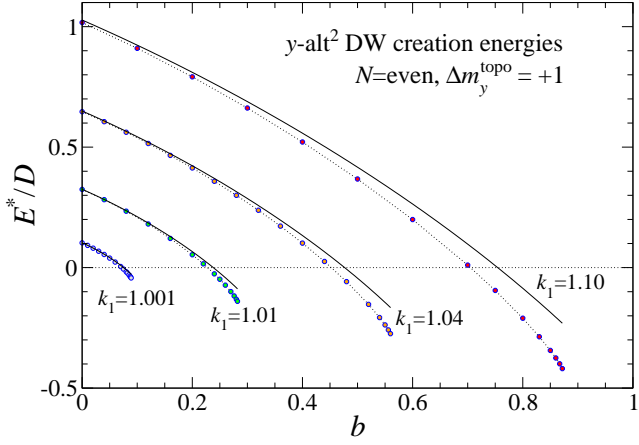}
\caption{\label{Eplus} Creation energy  of a $y$-alt$^2$ DW in applied magnetic field $b$, for various 
anisotropy strengths, even $N$, and $m_y^{\rm topo}(\text{DW})=+1$ in the DW state, or $\Delta m_y^{\rm topo}=+1$.  
Points are from simulations of chains with 500 or 800 sites, dotted lines are guides to the eye, and the 
solid black curves are the theory expression in Eq.\ (\ref{E*}), which accounts for the topological Oersted 
energy shift of the DW state.}
\end{figure}

When a transverse field is present, the continuum energy density in a static DW from Eq.\ (\ref{H-or}) with
$\Theta=\vartheta=0$ now includes terms in the small angle $\varphi$, 
\begin{align}
\frac{\cal H}{D} & = -1-\cos^2\!\Phi+({1-\tfrac{1}{2}\cos^2\!\Phi})\Phi_x^2 +3\varphi^2 
\nonumber \\
& -k_1 (1-\cos^2\!\Phi+\varphi^2 \cos 2\Phi) -b\varphi \cos\Phi.
\end{align}
This accounts for the induced magnetic moment, but not an intrinsic one. Using the asymptotic 
DW solution, Eqs.\ (\ref{psi-b}) and (\ref{vphi+b}), this can be written
\begin{align}
\frac{\cal H}{D} & = 
 -(1+k_1) +(k_1-1) \sin^2\!\psi 
\nonumber \\
& +(1-\tfrac{1}{2}\sin^2\psi)\psi_x^2 
 -\frac{b^2 \sin^2\!\psi}{4(3+k_1 \cos 2\psi)}.
\end{align}
The first term is the uniform $y$-alt energy density, which is to be subtracted out to get the DW creation energy.
The Oersted field term can benefit from an expansion for $\sin\psi\ll 1$,
\be
\frac{1}{3+k_1\cos 2\psi} \approx \frac{1}{3+k_1}\Big(1+\frac{2k_1}{3+k_1}\sin^2\!\psi\Big).
\ee
Then the continuum DW creation energy is an integral of the energy density over the system, 
\begin{align}
\frac{E_{\rm con}^*}{D} = & \int_{-\infty}^{\infty} dx
\Big\{ (k_1-1)\sin^2\!\psi+(1-\tfrac{1}{2}\sin^2\!\psi)\psi_x^2 
\nonumber \\
& -\frac{b^2 \sin^2\!\psi}{4(3+k_1)}\Big(1+\frac{2k_1}{3+k_1}\sin^2\!\psi\Big)\Big\}.
\end{align}
With a substitution based on Eq.\ (\ref{psi-b}), 
\be
k_1-1 = \beta^2 +\frac{b^2}{4(3+k_1)},
\ee
the integrals of $\sin^2\!\psi$ proportional to $b^2$ cancel out.  
The dipolar parts have been integrated earlier in Eq.\ (\ref{int123}).
The field contribution now depends on an integral evaluated numerically,
\be
\label{int4}
\int_{-\infty}^{\infty} dx\ \sin^4\!\psi  
\approx 1.03886429\ \beta^{-1}.
\ee
Combining all contributions, we see that the preliminary formula for continuum creation energy is the zero-field formula 
(\ref{EDW}), reduced by an Oersted term quadratic in the field,
\be
\label{Econ*}
\frac{E_{\rm con}^{*}}{D} \approx 3.2488\, \beta -1.03886 \frac{k_1 b^2}{2(3+k_1)^2\beta}.
\ee 
The inverse half-width $\beta$ also diminishes with increasing $b^2$. This preliminary formula only involves the 
induced magnetic moment $m_y$ and not any topological magnetic moment.  

\begin{figure}
\includegraphics[width=\figwidth,angle=0]{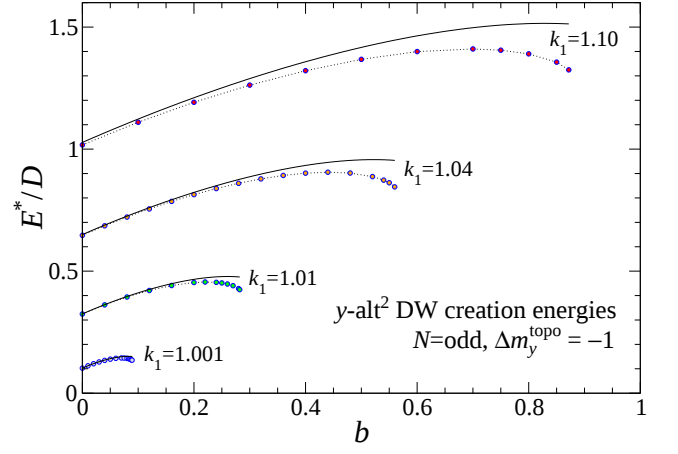}
\caption{\label{Eminus} Creation energy  of a $y$-alt$^2$ DW in applied magnetic field $b$, for various
anisotropy strengths, odd $N$, and $m_y^{\rm topo}(y\text{-alt})=+1$ in the reference $y$-alt state, or 
$\Delta m_y^{\rm topo} = -1$.  Points are from simulations of chains with 501 or 801 sites, dotted lines are 
guides to the eye, and the solid black curves are the theory expression in Eq.\ (\ref{E*}), which accounts for the 
topological Oersted energy shift of the odd-chain $y$-alt reference state.}
\end{figure}

To compare with $E^*$ in simulations, some care is needed, due to the topological magnetic moment for
DWs in even-$N$ chains and not in odd-$N$ chains.  The creation energy is the difference of the system energy 
with the DW present, minus the system energy in a uniform $y$-alt state, accounting correctly for 
intrinsic magnetic moments.  

If $N$ is even, the DW state has the induced field whose Oersted energy is included in Eq.\ (\ref{Econ*})
as a term dependent on $b^2$, plus a contribution due to the topological magnetic moment, 
$m_y^{\rm topo}(\text{DW})=\pm 1$.  The contribution of a topological or intrinsic magnetic moment has 
been left out of Eq.\ (\ref{Econ*}).  The extra topological Oersted energy contribution is 
$-b m_y^{\rm topo}(\text{DW})$.  Also note, the uniform $y$-alt state for even $N$ has no 
topological magnetic moment, as all spins are in up-down pairs.  

If $N$ is odd, the DW state has the induced field whose Oersted energy is included in Eq.\ (\ref{Econ*}), but
there is no topological magnetic moment.  However, for an odd chain, the $y$-alt state whose energy must be 
subtracted does have one unmatched spin, either up or down along $y$, giving {\em it} a topological magnetic moment
$m_y^{\rm topo}(y\text{-alt})= \pm 1$.  The corresponding energy contribution to the $y$-alt state is 
$-b m_y^{\rm topo}(y\text{-alt})$, which is to be subtracted out from $E_{\rm con}$.

Then the net creation energy for both even-$N$ and odd-$N$ chains can be written with a single formula,
\be
\label{E*}
\frac{E^{*}}{D} \approx 3.2488\, \beta -1.03886 \frac{k_1 b^2}{2(3+k_1)^2\beta}-b \Delta m_y^{\rm topo}.
\ee 
The topological term is expressed with the {\em change} in topological magnetic moment upon DW formation out of the 
$y$-alt state, defined as
\be
\Delta m_y^{\rm topo} \equiv m_y^{\rm topo}(\text{DW})-m_y^{\rm topo}(y\text{-alt}) .
\ee
Physically, that is the intrinsic magnetic moment carried into the system by the DW, if one imagined it
entering at one end of a long chain initially in a $y$-alt state. 
It can't enter any other way without tearing the texture. 
The creation energy formula predicts a combination of linear and roughly quadratic dependencies on $b$, 
with strongly different slopes as determined by $\Delta m_y^{\rm topo} = \pm 1$. 

Fig.\ \ref{Eplus} shows a comparison of the creation energies from simulations with this theoretical expression 
for $\Delta m_y^{\rm topo} = +1$, which means even chains where the DW has $m_y^{\rm topo}(\text{DW})=+1$ or odd chains
where the initial state has $m_y^{\rm topo}(y\text{-alt})= -1$.   Note how $E^{*}$ has
a noticeable negative slope, due to the topological term, whose explanation is elusive without
realizing that the continuum theory only includes the part of $m^y$ induced by the field. The simulated energies
are less than the theoretical values, as the configuration can deform away from the theoretical shape to better 
minimize its energy. Indeed, at large enough $b$, the creation energies become negative, and the DW state
has a lower energy than the uniform $y$-alt state.

Further results are seen in Fig.\ \ref{Eminus} for $\Delta m_y^{\rm topo} = -1$, which means even chains where
the DW has $m_y^{\rm topo}(\text{DW})=-1$ or odd chains where the initial state has $m_y^{\rm topo}(y\text{-alt})= +1$. 
Now the creation energy vs.\ $b$ initially has a positive slope due to the topological term.
The creation energy in this case does not become negative for any of the examples used.  For both signs of
$\Delta m_y^{\rm topo}$, the theory works best for values of $k_1$ as close as possible to unity, where 
a continuum view of the system is best (wide DWs).

\begin{figure}
\includegraphics[width=\figwidth,angle=0]{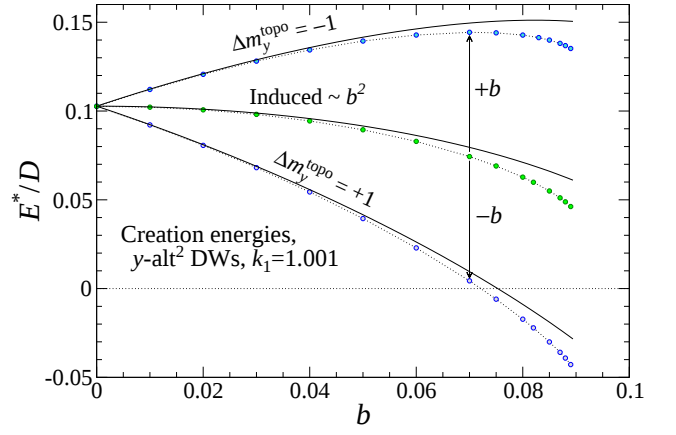}
\caption{\label{Edw1001b} Creation energy of $y$-alt$^2$ DWs vs.\ applied magnetic field $b$, 
for $\Delta m_y^{\rm topo}=\pm 1$ at a selected anisotropy.  Points are from simulations with 
$N=501$ and $\Delta m_y^{\rm topo}=-1$ (upper) 
and with $N=500$ and $\Delta m_y^{\rm topo}=+1$ (lower). 
Dotted lines are guides to the eye, and the solid black curves are the theory expressions
in Eq.\ (\ref{E*}), which include the topological Oersted energy shifts, indicated by arrows.
The intermediate data and curve show simulations and theoretical Eq.\ (\ref{Econ*}) that contain
the energy contributions due to induced magnetic moment without topological effects.}
\end{figure}

It is also interesting to directly compare the creation energies for $\Delta m_y^{\rm topo}=\pm 1$. At $k_1=1.001$,
that is shown in Fig.\ \ref{Edw1001b}.  The curves start at the same point for $b=0$ and then separate by a difference 
of $2b$ for increasing field. Each curve is shifted away from an average curve (the energy of the induced magnetic moment)
described by Eq.\ (\ref{Econ*}) by $\pm b$ due to the topological effects.

\section{Discussion and Conclusions}
The 1D chain of transverse magnetic islands analyzed here has been shown to have pairs of degenerate
uniform states, between which domain walls can form.  For the pair of almost indistinguishable  
uniform $y$-alt states, such as those present at opposite ends in Fig.\ \ref{oval-islands-yaltdw},
the DW imposes a rotation of the dipoles on alternating sites in opposing senses in the $xy$
plane.  This was found initially via numerical simulations that relax the system from an initial 
abrupt DW towards one that is spread out and a local energy minimum.  In static equilibrium, the dipoles 
lie in the $xy$ plane, each parallel to the effective field ${\bf B}_n$ in Eq.\ (\ref{dyn1}) that acts on it. 

Due to that rotation within the DW,  a large net magnetic moment $m_x$ parallel to the chain emerges under 
this opposing rotations scenario. Its magnitude in spin units is of the order of the DW full-width,
as estimated in Eq.\ (\ref{mxa}).  If there also happens to be an applied magnetic field transverse to 
the chain, that also induces a weaker transverse magnetic moment $m_y$, whose magnitude increases nonlinearly
with the field strength, as in Fig.\ \ref{myodd}.

For the simplest continuum analysis, NN dipolar interactions together with island anisotropies have been considered.  
A set of continuum limit equations of motion for the dipolar angles has been found after tedious calculations in
two different ways for verification: (1) expansion of the discrete dynamic equations, and (2) by derivation from 
a continuum Hamiltonian.  These derivations used the assumption of opposing rotations on the two sublattices,
which is in contrast to an opposing directions assumption that is traditionally used for antiferromagnets.
This difference is critical in obtaining a continuum description for NN dipolar interactions, which energetically
favor, in a local sense,  both NN spins antiparallel and transverse to the chain, or NN spins parallel to the chain.   
Due to these physical effects, the spins in a $y$-alt$^2$ DW indeed follow a rotation path that keeps neighboring
spins close to minimizing the dipolar energies, while they compete with weak anisotropy and Oersted energies.

The static DW equilibrium equation (\ref{EE}) for large in-plane angle $\Phi$ has strong nonlinearity, however, it has 
been approximated and a linear equation (\ref{linear}) holds for the deviation $\psi$ of the spins' directions away
from perpendicular to the chain. The approximate solution (\ref{DW}) is surprisingly good at capturing the $y$-alt$^2$ DW   
structure, even in its center. A partial linearization instead leads to a sine-Gordon equation, whose solution in Eq.\ 
(\ref{sGkink}) does not follow the DW structure found in numerical relaxation. 

For zero applied field, a $y$-alt$^2$ DW's transverse magnetic moment $m_y$ is zero for a DW in chains
with an odd number of sites, but takes only the discrete values $m_y =\pm 1$ (in spin units) for chains with
an even number of sites.  For odd chains, the central site can be isolated and contributes zero, while
the two remaining halves of the system have equal and opposite contributions to $m_y$, netting out zero.
Adding one more spin at an end, either up or down along $y$, then nets out $m_y =\pm 1$ in even chains.  An 
analysis was given to interpret this as a topological magnetic moment of the DW itself, which also directly
affects the creation energy of the DWs, especially in an applied field.  

Although not analyzed here, one can ask how the results could change when long-range dipolar (LRD) interactions are
included.  For the NN model, we noticed that $k_1=1$ is a type of critical point where the DW width diverges,
and the uniform $y$-alt states it connects are destabilized into $x$-parallel states when $b=0$ or oblique states 
when $b\ne 0$.  Including LRD interactions shifts this instability point for uniform states to 
$k_1 \approx 1.50257$, a result found both from numerics and continuum theory \cite{Wysin24}.  It is expected 
that a similar shift of the critical point for $y$-alt$^2$ DWs will happen.  

Currently, a reliable way to include longer-range dipole
interactions in the continuum theory while under the opposing rotations assumptions has not been developed.
The reason is that the continuum limit (smooth NN variations) and LRD interactions (discrete NN variations) are 
incompatible.  Continuum theory for DWs found in numerics that connect oblique states to $y$-alt states also 
offers an unsolved challenge.  Part of the motivation for this report is to stimulate further research in related 
condensed matter models with DWs of alternating order.

\section{Funding}
This research received no external funding.

\section{Data Availability Statement}
All data relevant to this research is included in the article.

\section{Conflicts of Interest}
The author declares no conflicts of interest.

\begin{widetext}
\appendix
\section{Expansion of the discrete equations of motion}
\label{App1}
For the continuum expansion of the dynamic equations (\ref{dotphi}) and (\ref{dottheta}) there are contributions
from the dipolar interactions, anisotropy, and applied field.  They can be derived starting from 
the expansions for cosine and sine of $\phi_{n+1}$ (or $\theta_{n+1}$) given in Eqs.\ (\ref{cosa}) and (\ref{sina}).
Dot indicates derivative with respect to time and lattice constant $a$ will be set to 1 at the end, 
which implies continuum position $x$ is measured in units of $a$.

\subsection{Dynamics of the in-plane angles}
The first dipolar contribution to $\dot\phi_n$ for an even-$n$ site is a sum over right and left neighbors, 
indicated by $\sum_{\pm a}$, 
\begin{align}
d_1 & \equiv \sin\theta_{n-1}+\sin\theta_{n+1}
 \approx \sum_{\pm a} \big\{ -\sin\Theta[1-\tfrac{1}{2}\left(\vartheta^2+a^2\Theta_x^2-2a\vartheta\Theta_x\right)]
 +\cos\Theta[\vartheta-a\Theta_x-\tfrac{1}{2}a^2\Theta_{xx}] \big\}
\nonumber \\
& \approx -\sin\Theta (2-\vartheta^2-a^2\Theta_x^2) +\cos\Theta (2\vartheta-a^2\Theta_{xx}).
\end{align}
That produces just twice the even terms in lattice constant $a$. 

The next dipole contribution to $\dot\phi_n$, in the 2nd line of Eq.\ (\ref{dotphi}), depends on a combination
of cosines of $\theta_{n+1}$ and $\phi_{n+1}$, followed by summing over nearest neighbors. 
We already know the individual expansion of $\cos\phi_{n+1}$ in Eq.\ (\ref{cosa}), and there is one 
of the same form for the out-of-plane angle, 
\begin{align}
\label{cosb}
\cos\theta_{n+1} 
& \approx \cos\Theta [1-\tfrac{1}{2}\left(\vartheta^2+a^2\Theta_x^2-2a\vartheta\Theta_x\right)]
 +\sin\Theta [\vartheta-a\Theta_x-\tfrac{1}{2}a^2\Theta_{xx}].
\end{align}
Forming the products of cosines and summing over $n+1$ and $n-1$ terms, while keeping up to quadratic terms, leaves,
\begin{align}
\label{ctcp}
c_n & \equiv 
\cos\theta_{n-1}\cos\phi_{n-1}+\cos\theta_{n+1}\cos\phi_{n+1}
\\
& \approx \cos\Theta\cos\Phi[2-\{\vartheta^2+\varphi^2+a^2(\Theta_x^2+\Phi_x^2)\}]
 +\sin\Theta\sin\Phi[2\vartheta\varphi+2a^2\Theta_x\Phi_x]
\nonumber \\
& +\cos\Theta\sin\Phi[2\varphi-a^2\Phi_{xx}] +\sin\Theta\cos\Phi[2\vartheta-a^2\Theta_{xx}].
\nonumber
\end{align} 
To get its net dynamic contribution for an even-$n$ site, one multiplies $c_n$ by $2\tan\theta_n\cos\phi_n$, which
requires
\begin{align}
\tan(\Theta+\vartheta) 
 & \approx \tan\Theta+\vartheta \sec^2\!\Theta +\vartheta^2\sec^2\!\Theta\tan\Theta,
\nonumber \\
\cos(\Phi+\varphi) 
& \approx \left(1-\tfrac{1}{2}\varphi^2\right)\cos\Phi -\varphi\sin\Phi.
\end{align}
Combining all the factors and limiting to quadratic terms give the next contribution to $\dot\phi_n$ 
from dipole interactions,
\begin{align}
d_2 & \equiv 2\tan\theta_n \cos\phi_n (\cos\theta_{n-1}\cos\phi_{n-1}+\cos\theta_{n+1}\cos\phi_{n+1})
\nonumber \\
& \approx 2\left\{ \left(1-\tfrac{1}{2}\varphi^2\right) \tan\Theta\cos\Phi
-\varphi\tan\Theta\sin\Phi+\vartheta \sec^2\!\Theta\left(\cos\Phi
+\vartheta\tan\Theta\cos\Phi-\varphi\sin\Phi\right) \right\}
\nonumber \\
& \times \big\{ \cos\Theta\cos\Phi[2-\{\vartheta^2+\varphi^2+a^2(\Theta_x^2+\Phi_x^2)\}]
+\sin\Theta\sin\Phi[2\vartheta\varphi+2a^2\Theta_x\Phi_x]
\nonumber \\
& +\cos\Theta\sin\Phi[2\varphi-a^2\Phi_{xx}] +\sin\Theta\cos\Phi[2\vartheta-a^2\Theta_{xx}]
\big\}
\nonumber \\
& \approx  2\big\{
\sin\Theta\cos^2\!\Phi [2-a^2(\Theta_x^2+\Phi_x^2)]
 +\tan\Theta\sin\Theta \sin 2\Phi (a^2\Theta_x\Phi_x)
-\sin\Theta \sin 2\Phi (\tfrac{1}{2}a^2\Phi_{xx})
\nonumber \\
& -\tan\Theta\sin\Theta \cos^2\!\Phi (a^2\Theta_{xx})
-2\varphi^2 \sin\Theta
+2\vartheta \sec\Theta (1+\sin^2\!\Theta) \cos^2\!\Phi
+\vartheta^2 (4\sec^2\!\Theta-1) \sin\Theta\cos^2\!\Phi
\big\}.
\end{align}
The third line of Eq.\ (\ref{dotphi}) contains the third dipole-interaction contribution to $\dot\phi_n$. It 
depends on a combination of cosines of $\theta_{n+1}$ and sines of $\phi_{n+1}$, followed by summing over nearest neighbors,
\begin{align}
\label{ctsp}
s_n & \equiv 
\cos\theta_{n-1}\sin\phi_{n-1}+\cos\theta_{n+1}\sin\phi_{n+1}
\\
& \approx -\cos\Theta\sin\Phi[2-\{\varphi^2+\vartheta^2+a^2(\Theta_x^2+\Phi_x^2)\}]
 +\sin\Theta\cos\Phi[2\vartheta\varphi+2a^2\Theta_x\Phi_x]
\nonumber \\
& +\cos\Theta\cos\Phi[2\varphi-a^2\Phi_{xx}] -\sin\Theta\sin\Phi[2\vartheta-a^2\Theta_{xx}].
\nonumber 
\end{align}
To get its net dynamic contribution for an even-$n$ site, one multiplies $s_n$ by $-\tan\theta_n\sin\phi_n$,
which produces this last dipole interaction contribution,
\begin{align}
d_3 \equiv & -\tan\theta_n \sin\phi_n\ (\cos\theta_{n-k}\sin\phi_{n-k}+\cos\theta_{n+k}\sin\phi_{n+k})
\nonumber \\
& \approx -\left\{\left(1-\tfrac{1}{2}\varphi^2\right) \tan\Theta\sin\Phi
+\varphi\tan\Theta\cos\Phi
+\vartheta\sec^2\!\Theta (\sin\Phi+\vartheta\tan\Theta\sin\Phi+\varphi\cos\Phi) \right\}
\nonumber \\
& \times \big\{-\cos\Theta\sin\Phi[2-\{\varphi^2+\vartheta^2+a^2(\Theta_x^2+\Phi_x^2)\}]
+\sin\Theta\cos\Phi [2\vartheta\varphi+2a^2\Theta_x\Phi_x]
\nonumber \\
& +\cos\Theta\cos\Phi [2\varphi-a^2\Phi_{xx}]
-\sin\Theta\sin\Phi [2\vartheta-a^2\Theta_{xx}] 
\big\}
\nonumber \\
& \approx -\big\{ 
-\sin\Theta\sin^2\!\Phi [2-a^2(\Theta_x^2+\Phi_x^2)]
+\tan\Theta\sin\Theta \sin 2\Phi (a^2\Theta_x\Phi_x)
-\sin\Theta \sin 2\Phi (\tfrac{1}{2}a^2\Phi_{xx})
\nonumber \\
& +\tan\Theta\sin\Theta \sin^2\!\Phi (a^2\Theta_{xx}) +2\varphi^2 \sin\Theta
-2\vartheta \sec\Theta(1+\sin^2\!\Theta) \sin^2\!\Phi
-\vartheta^2(4\sec^2\!\Theta-1)\sin\Theta \sin^2\!\Phi
\big\}.
\end{align}
The expression inside the final braces can be obtained from the corresponding one for $d_2$ via the
symmetry changes $\cos^2\!\Phi \to\-\sin^2\!\Phi$, $\sin^2\!\Phi\to -\cos^2\!\Phi$, with 
$\sin 2\Phi = 2\sin\Phi \cos\Phi$ being unchanged.  The term with $\varphi^2\sin\Theta$ also reverses sign, 
because it came from the combination $\cos^2\!\Phi+\sin^2\!\Phi$, which reverses sign.

The combination of the three dipolar parts is the time derivative contribution from dipolar effects,
written for an even-$n$ site as
\begin{align}
\dot\phi_{\text{dip, evn}} & = (\dot\Phi+\dot\varphi)_{\rm dip} = d_1+d_2+d_3 
\nonumber \\
& \approx  
\sin\Theta \cos^2\!\Phi (2-a^2\Theta_x^2) -(1+\cos^2\!\Phi)\sin\Theta(a^2\Phi_x^2)
+\tan\Theta\sin\Theta \sin 2\Phi (a^2\Theta_x\Phi_x)
\nonumber \\
 &  -\sin\Theta \sin 2\Phi (\tfrac{1}{2}a^2\Phi_{xx})
-\sec\Theta(1+\sin^2\!\Theta\cos^2\!\Phi)a^2\Theta_{xx} -6\varphi^2 \sin\Theta
\nonumber \\
& +2\vartheta \sec\Theta\left[2+\left(1+\sin^2\!\Theta\right)\cos^2\!\Phi\right]
+4\vartheta^2 \sin\Theta \left[ \sec^2\!\Theta+\left(\sec^2\!\Theta-\tfrac{1}{4}\right)\cos^2\!\Phi \right] .
\end{align}
On the odd-$n$ sublattice under the opposing rotations assumption, however, the signs of $\Phi$ and
$\Theta$ get reversed, while $\varphi$ and $\vartheta$ are unchanged. Then it is easy to see that the
dipolar effects on an odd site can be written as 
\begin{align}
\dot\phi_{\text{dip, odd}} & = (-\dot\Phi+\dot\varphi)_{\rm dip} = d_1+d_2+d_3
\nonumber \\
& \approx
-\sin\Theta \cos^2\!\Phi (2-a^2\Theta_x^2) +(1+\cos^2\!\Phi)\sin\Theta(a^2\Phi_x^2)
-\tan\Theta\sin\Theta \sin 2\Phi (a^2\Theta_x\Phi_x)
\nonumber \\
 &  +\sin\Theta \sin 2\Phi (\tfrac{1}{2}a^2\Phi_{xx})
+\sec\Theta(1+\sin^2\!\Theta\cos^2\!\Phi)a^2\Theta_{xx} +6\varphi^2 \sin\Theta
\nonumber \\
& +2\vartheta \sec\Theta\left[2+\left(1+\sin^2\!\Theta\right)\cos^2\!\Phi\right]
-4\vartheta^2 \sin\Theta \left[ \sec^2\!\Theta+\left(\sec^2\!\Theta-\tfrac{1}{4}\right)\cos^2\!\Phi \right] .
\end{align}
Following similar but simpler procedures, the contributions from anisotropy for the even-$n$ sublattice 
in the 4th line of Eq.\ (\ref{dotphi}) are 
\begin{align}
\dot\phi_{\text{ani, evn}}\ & = (\dot\Phi+\dot\varphi)_{\rm ani} 
= \sin(\Theta+\vartheta) \left[2k_1 \sin^2\!(\Phi+\varphi) +2k_3\right]
\nonumber \\
& \approx 2
\left\{\left(1-\tfrac{1}{2}\vartheta^2\right)\sin\Theta+\vartheta \cos\Theta \right\}
\left\{ k_1 [\sin^2\!\Phi +\varphi\sin 2\Phi + \varphi^2 \cos 2\Phi ] +k_3 \right\}. 
\end{align}
For a site on the odd sublattice, the signs of the large angles are reversed,
\begin{align}
\dot\phi_{\text{ani, odd}}\ & = (-\dot\Phi+\dot\varphi)_{\rm ani}
= \sin(-\Theta+\vartheta) \left[2k_1 \sin^2\!(-\Phi+\varphi) +2k_3\right]
\nonumber \\
& \approx 2
\left\{-\left(1-\tfrac{1}{2}\vartheta^2\right)\sin\Theta+\vartheta \cos\Theta \right\}
\left\{ k_1 [\sin^2\!\Phi -\varphi\sin 2\Phi + \varphi^2 \cos 2\Phi ] +k_3 \right\}.
\end{align}
Finally, there are the applied field contributions, first for an even-$n$ site,
\begin{align}
\dot\phi_{\text{field, evn}}\ & = (\dot\Phi+\dot\varphi)_{\rm field} = b \tan(\Theta+\vartheta)\sin(\Phi+\varphi)
\nonumber \\
& \approx b (\tan\Theta+\vartheta\sec^2\!\Theta+\vartheta^2\sec^2\!\Theta\tan\Theta)
\left[ \left(1-\tfrac{1}{2}\varphi^2\right)\sin\Phi+\varphi\cos\Phi \right]
\nonumber \\
& \approx b \big[ \left(1-\tfrac{1}{2}\varphi^2\right)\tan\Theta\sin\Phi
+\vartheta\sec^2\!\Theta(\sin\Phi+\varphi\cos\Phi)
+\varphi \tan\Theta\cos\Phi +\vartheta^2\sec^2\!\Theta\tan\Theta\sin\Phi\big].
\end{align}
Reverse the signs of the large angles for an odd-$n$ site,
\begin{align}
\dot\phi_{\text{field, odd}}\ & = (-\dot\Phi+\dot\varphi)_{\rm field} = b \tan(-\Theta+\vartheta)\sin(-\Phi+\varphi)
\nonumber \\
& \approx b (-\tan\Theta-\vartheta\sec^2\!\Theta+\vartheta^2\sec^2\!\Theta\tan\Theta)
\left[ -\left(1-\tfrac{1}{2}\varphi^2\right)\sin\Phi+\varphi\cos\Phi \right]
\nonumber \\
& \approx b \big[ \left(1-\tfrac{1}{2}\varphi^2\right)\tan\Theta\sin\Phi
+\vartheta\sec^2\!\Theta(-\sin\Phi+\varphi\cos\Phi)
-\varphi \tan\Theta\cos\Phi +\vartheta^2\sec^2\!\Theta\tan\Theta\sin\Phi\big].
\end{align}

Based on the above results, the time derivatives of the in-plane continuum angles are related to those for
the sublattice angles by
\be
\begin{array}{l}
\dot\phi_{\rm evn} = +\dot\Phi+\dot\varphi \\ \\
\dot\phi_{\rm odd} = -\dot\Phi+\dot\varphi
\end{array}
\hskip 0.5cm \implies \hskip 0.5cm
\begin{array}{l}
\dot\Phi = \tfrac{1}{2}(\dot\phi_{\rm evn}-\dot\phi_{\rm odd})  \\  \\
\dot\varphi = \tfrac{1}{2}(\dot\phi_{\rm evn}+\dot\phi_{\rm odd}).
\end{array}
\ee
Since the odd formulas are obtained from the evens by reversing the large angles, the terms of odd symmetry 
in $\Theta,\Phi$ contribute to $\dot\Phi$ and those of even symmetry contribute to $\dot\varphi$.
What results for the large in-plane angle is:
\begin{align}
\label{dPhidt}
\dot\Phi & = \tfrac{1}{2}(\dot\phi_{\rm evn}-\dot\phi_{\rm odd})
= \tfrac{1}{2}[\dot\phi_{\rm evn}(\Theta,\Phi)-\dot\phi_{\rm evn}(-\Theta,-\Phi)]
\nonumber \\
& \approx 
\sin\Theta \cos^2\!\Phi (2-\Theta_x^2) -(1+\cos^2\!\Phi)\sin\Theta(\Phi_x^2)
+\tan\Theta\sin\Theta \sin 2\Phi (\Theta_x\Phi_x)
\nonumber \\
 &  -\sin\Theta \sin 2\Phi (\tfrac{1}{2}\Phi_{xx})
-\sec\Theta(1+\sin^2\!\Theta\cos^2\!\Phi)\Theta_{xx}  -6\varphi^2 \sin\Theta
\nonumber \\
& +4\vartheta^2 \sin\Theta \left[ \sec^2\!\Theta+\left(\sec^2\!\Theta-\tfrac{1}{4}\right)\cos^2\!\Phi \right]
\nonumber \\
& +2k_1 \left[\left(1-\tfrac{1}{2}\vartheta^2\right)\sin\Theta \sin^2\!\Phi
+\varphi^2 \sin\Theta\cos 2\Phi+\varphi\vartheta\cos\Theta\sin 2\Phi \right]
+2k_3 \left(1-\tfrac{1}{2}\vartheta^2\right)\sin\Theta
\nonumber \\
& +b \left( \vartheta\sec^2\!\Theta\sin\Phi+\varphi \tan\Theta\cos\Phi \right) .
\end{align}
Similarly, for the small in-plane angle, the even-symmetry terms give:
\begin{align}
\label{dphidt}
\dot\varphi & = \tfrac{1}{2}(\dot\phi_{\rm evn}+\dot\phi_{\rm odd})
= \tfrac{1}{2}[\dot\phi_{\rm evn}(\Theta,\Phi)+\dot\phi_{\rm evn}(-\Theta,-\Phi)]
\nonumber \\
& \approx 2\vartheta \sec\Theta\left\{2+\left(1+\sin^2\!\Theta\right)\cos^2\!\Phi\right\}
+2k_1 \left(\vartheta\cos\Theta\sin^2\!\Phi+\varphi\sin\Theta\sin 2\Phi \right)
+2k_3 \vartheta \cos\Theta
\nonumber \\
& +b\left[ \left(1-\tfrac{1}{2}\varphi^2+\vartheta^2\sec^2\!\Theta\right)\tan\Theta\sin\Phi
+\vartheta\varphi\sec^2\!\Theta\cos\Phi \right]. 
\end{align}
In both equations, all terms up to quadratic order in space derivatives and small angles have been kept.

\subsection{Dynamics of the out-of-plane angles}
For the continuum expansion of $\dot\theta_n$ using Eq.\ (\ref{dottheta}), one sees that the dipolar
parts again contain the factors $c_n$ and $s_n$ defined above. They are multiplied by $-2\sin\phi_n$ 
and $-\cos\phi_n$, respectively.  This gives the first dipolar contribution for an even-$n$ site,
\begin{align}
d_1 & \equiv -(2\sin\phi_n) c_n
= -2\sin(\Phi+\varphi) (\cos\theta_{n-1}\cos\phi_{n-1}+\cos\theta_{n+1}\cos\phi_{n+1})
\nonumber \\
& \approx -2\big[(1-\tfrac{1}{2}\varphi^2) \sin\Phi +\varphi \cos\Phi \big]
\times 
\big\{ \cos\Theta\cos\Phi[2-\{\vartheta^2+\varphi^2+(\Theta_x^2+\Phi_x^2)\}]
\nonumber \\
& +\sin\Theta\sin\Phi(2\vartheta\varphi+2\Theta_x\Phi_x)
 +\cos\Theta\sin\Phi(2\varphi-\Phi_{xx}) +\sin\Theta\cos\Phi(2\vartheta-\Theta_{xx})
\big\}
\nonumber \\
& \approx -2\big\{ 
\cos\Theta \big[ \sin 2\Phi +2\varphi 
-\tfrac{1}{2}\sin 2\Phi(\vartheta^2+\Theta_x^2+\Phi_x^2)-\sin^2\!\Phi(\Phi_{xx}) \big]
\nonumber \\
& +\sin\Theta \big[ 
\sin 2\Phi (\vartheta-\tfrac{1}{2}\Theta_{xx}) +2\vartheta\varphi +\sin^2\!\Phi(2\Theta_x\Phi_x) 
\big]
\big\}.
\end{align}
The second contribution for an even-$n$ site is
\begin{align}
d_2 & \equiv -(\cos\phi_n) s_n
= -\cos(\Phi+\varphi) (\cos\theta_{n-1}\sin\phi_{n-1}+\cos\theta_{n+1}\sin\phi_{n+1})
\nonumber \\
& \approx -\big[(1-\tfrac{1}{2}\varphi^2) \cos\Phi -\varphi \sin\Phi \big]
\times 
\big\{-\cos\Theta\sin\Phi[2-\{\varphi^2+\vartheta^2+(\Theta_x^2+\Phi_x^2)\}]
\nonumber \\
& +\sin\Theta\cos\Phi(2\vartheta\varphi+2\Theta_x\Phi_x)
 +\cos\Theta\cos\Phi(2\varphi-\Phi_{xx})
-\sin\Theta\sin\Phi(2\vartheta-\Theta_{xx})
\big\}
\nonumber \\
& \approx -\big\{ 
\cos\Theta \big[ -\sin 2\Phi +2\varphi 
+\tfrac{1}{2}\sin 2\Phi (\vartheta^2+\Theta_x^2+\Phi_x^2) -\cos^2\!\Phi(\Phi_{xx}) \big]
\nonumber \\
& +\sin\Theta \big[ 
-\sin 2\Phi (\vartheta-\tfrac{1}{2}\Theta_{xx}) +2\vartheta\varphi +\cos^2\!\Phi (2\Theta_x\Phi_x) \big]
\big\} .
\end{align}
The two factors $d_1,\ d_2$ are very similar; note the swaps of $\sin^2\!\Phi$ with $\cos^2\!\Phi$ and
the sign changes on the $\sin 2\Phi$ terms.  Combined, these give the dipolar contribution to the 
time-derivative on an even-$n$ site,
\begin{align}
\dot\theta_{\text{dip, evn}} & = (\Theta+\vartheta)_{\rm dip} = d_1+d_2
\nonumber \\
& \approx -\big\{
\cos\Theta \big[ \sin 2\Phi +6\varphi
-\tfrac{1}{2}\sin 2\Phi (\vartheta^2+\Theta_x^2+\Phi_x^2) -(2\sin^2\!\Phi+\cos^2\!\Phi)(\Phi_{xx}) \big]
\nonumber \\
& +\sin\Theta \big[
\sin 2\Phi (\vartheta-\tfrac{1}{2}\Theta_{xx}) +6\vartheta\varphi +(2\sin^2\!\Phi+\cos^2\!\Phi) (2\Theta_x\Phi_x) \big]
\big\} .
\end{align}
Similarly, for odd-$n$ sites under the opposing rotations assumption, $\Phi$ and $\Theta$ reverse signs, which
gives, 
\begin{align}
\dot\theta_{\text{dip, odd}} & = (-\Theta+\vartheta)_{\rm dip} = d_1+d_2
\nonumber \\
& \approx -\big\{
\cos\Theta \big[ -\sin 2\Phi +6\varphi
+\tfrac{1}{2}\sin 2\Phi (\vartheta^2+\Theta_x^2+\Phi_x^2) +(2\sin^2\!\Phi+\cos^2\!\Phi)(\Phi_{xx}) \big]
\nonumber \\
& +\sin\Theta \big[
\sin 2\Phi (\vartheta+\tfrac{1}{2}\Theta_{xx}) -6\vartheta\varphi -(2\sin^2\!\Phi+\cos^2\!\Phi) (2\Theta_x\Phi_x) \big]
\big\} .
\end{align}
The contributions from anisotropy for the even-$n$ sites in Eq.\ (\ref{dottheta}) are
\begin{align}
\dot\theta_{\text{ani, evn}} 
& = (\dot\Theta+\dot\vartheta)_{\rm ani} = k_1 \cos(\Theta+\vartheta) \sin 2(\Phi+\varphi)
\nonumber \\
& \approx k_1 [(1-\tfrac{1}{2}\vartheta^2)\cos\Theta-\vartheta \sin\Theta]
[(1-2\varphi^2)\sin 2\Phi+2\varphi \cos 2\Phi]
\nonumber \\
& \approx  k_1 \big\{ \cos\Theta [(1-2\varphi^2-\tfrac{1}{2}\vartheta^2)\sin 2\Phi
+2\varphi \cos 2\Phi] +\sin\Theta [-\vartheta\sin 2\Phi-2\vartheta\varphi \cos 2\Phi] \big\}.
\end{align}
For the odd-$n$ sites, reverse the large angles, to give,
\begin{align}
\dot\theta_{\text{ani, odd}}
& = (-\dot\Theta+\dot\vartheta)_{\rm ani} = k_1 \cos(-\Theta+\vartheta) \sin 2(-\Phi+\varphi)
\nonumber \\
& \approx k_1 [(1-\tfrac{1}{2}\vartheta^2)\cos\Theta+\vartheta \sin\Theta]
[-(1-2\varphi^2)\sin 2\Phi+2\varphi \cos 2\Phi]
\nonumber \\
& \approx  k_1 \big\{ \cos\Theta [-(1-2\varphi^2-\tfrac{1}{2}\vartheta^2)\sin 2\Phi
+2\varphi \cos 2\Phi] +\sin\Theta [-\vartheta\sin 2\Phi+2\vartheta\varphi \cos 2\Phi] \big\}.
\end{align}
There are also the applied field contributions, found similarly,
\begin{align}
\dot\theta_{\text{field, evn}}
& = (+\dot\Theta+\dot\vartheta)_{\rm field} = b \cos(+\Phi+\varphi)
\approx b [(1-\tfrac{1}{2}\varphi^2)\cos\Phi-\varphi\sin\Phi],
\nonumber \\
\dot\theta_{\text{field, odd}}
& = (-\dot\Theta+\dot\vartheta)_{\rm field} = b \cos(-\Phi+\varphi)
\approx b [(1-\tfrac{1}{2}\varphi^2)\cos\Phi+\varphi\sin\Phi].
\end{align}
As for in-plane angles, the connection from discrete to continuum angles is
\be
\begin{array}{l}
\dot\theta_{\rm evn} = +\dot\Theta+\dot\vartheta \\ \\
\dot\theta_{\rm odd} = -\dot\Theta+\dot\vartheta
\end{array}
\hskip 1cm \implies \hskip 1cm
\begin{array}{l}
\dot\Theta = \tfrac{1}{2}(\dot\theta_{\rm evn}-\dot\theta_{\rm odd})  \\  \\
\dot\vartheta = \tfrac{1}{2}(\dot\theta_{\rm evn}+\dot\theta_{\rm odd}).
\end{array}
\ee
Then, the terms of odd symmetry in $\Theta,\Phi$ in $\dot\theta_{\rm evn}$ contribute to $\dot\Theta$, and
the terms of even symmetry contribute to $\dot\vartheta$.  Combining dipole, anisotropy, and field contributions, 
the dynamics for the large out-of-plane angle is then described by
\begin{align}
\label{dThetadt}
\dot\Theta & = \tfrac{1}{2}(\dot\theta_{\rm evn}-\dot\theta_{\rm odd}) =
\tfrac{1}{2}[\dot\theta_{\rm evn}(\Theta,\Phi)-\dot\theta_{\rm evn}(-\Theta,-\Phi)] 
\nonumber \\
& \approx \cos\Theta [(-1+\tfrac{1}{2}\{\vartheta^2+\Theta_x^2+\Phi_x^2\}) \sin 2\Phi
+(2\sin^2\!\Phi+\cos^2\!\Phi)\Phi_{xx} ]
\nonumber \\
& +\sin\Theta [(\tfrac{1}{2}\sin 2\Phi) \Theta_{xx} -6\vartheta\varphi
-(2\sin^2\!\Phi+\cos^2\!\Phi)(2\Theta_x\Phi_x)]
\nonumber \\
& +k_1 \big[ (1-2\varphi^2-\tfrac{1}{2}\vartheta^2)\cos\Theta\sin 2\Phi
-2\vartheta\varphi\sin\Theta\cos 2\Phi \big] -b\varphi\sin\Phi.
\end{align}
For the small out-of-plane angle, the remaining even-symmetry terms give
\begin{align}
\label{dthetadt}
\dot\vartheta & = \tfrac{1}{2}(\dot\theta_{\rm evn}+\dot\theta_{\rm odd}) =
\tfrac{1}{2}[\dot\theta_{\rm evn}(\Theta,\Phi)+\dot\theta_{\rm odd}(-\Theta,-\Phi)] 
\nonumber \\ 
& \approx -6\varphi \cos\Theta -\vartheta\sin\Theta \sin 2\Phi
 +k_1 ( 2\varphi \cos\Theta \cos 2\Phi -\vartheta\sin\Theta \sin 2 \Phi)
+b\left(1-\tfrac{1}{2}\varphi^2\right)\cos\Phi .
\end{align}

\section{Opposing rotations continuum dynamics with a Hamiltonian approach}
\label{App2}
The continuum dynamic equations under the assumption of opposing rotations on the two sublattices can be
derived starting from a general Hamiltonian for the system.  That Hamiltonian should already be known 
as a functional of the continuum in-plane angles $\Phi,\ \varphi$ and out-of-plane angles $\Theta,\ \theta$, and
their space derivatives. The goal here is to express equations (\ref{HamEq}) fully in terms of the continuum angles. 
This approach is complementary to the alternative of expanding the discrete equations of motion 
(\ref{dotphi}) and (\ref{dottheta}) involving the discrete angles $(\phi_n, \theta_n)$. 

\subsection{The continuum Hamiltonian}
The opposing rotations assumption in Eq.\ (\ref{map}) maps the discrete angles into the continuum angles
differently on the even and odd sublattices. The full Hamiltonian is a sum over sublattice contributions. To get the 
continuum density ${\cal H}(\Phi,\varphi,\Theta,\vartheta)$ for some $x$, an average is made over the energy
contribution from a site $n$, allowing it could be either an odd or even site.  The net $H$ can be expressed
as an average of partial Hamiltonians,
\be
H = \tfrac{1}{2} \left(H_{\rm evn}+H_{\rm odd}\right) 
\equiv  \sum_n \tfrac{1}{2}\left( H_{n(\text{evn})}+H_{n(\text{odd})}\right).
\ee
The expression in the sum will be converted into ${\cal H}(\Phi,\varphi,\Theta,\vartheta)$ with the different  
opposing rotations mappings to the continuum angles on the two sublattices. 

For the transverse island model limited to NN interactions, the discrete Hamiltonian from Eq.\ (\ref{Hn}) is
\begin{align}
\label{HNN}
H = \sum_{n} & \Big\{ 
D \left[ -2S_n^x S_{n+1}^x +S_n^y S_{n+1}^y +S_n^z S_{n+1}^z \right]
\nonumber \\
& -K_1 \left(S_n^y\right)^2 +K_3 \left(S_n^z\right)^2 -\mu B S_n^y \Big\}.
\end{align}
To convert to the continuum $H$, I will average over assuming $n=$even with $n=$odd. 

First consider the dipolar part involving $z$-components, using the opposing rotations assumption (\ref{map}),  
If $n=$even and $n+1=$odd, we have 
\begin{align}
\label{dzz}
S_n^z S_{n+1}^z & = \sin\theta_n \sin\theta_{n+1} = \sin[\Theta(x)+\vartheta] \sin[-\Theta(x+a)+\vartheta]
\nonumber \\
& = (\sin\Theta \cos\vartheta+\cos\Theta\sin\vartheta) (-)\sin(\Theta+a\Theta_x+\tfrac{1}{2}a^2\Theta_{xx}-\vartheta)
\nonumber \\
& \approx -\left[\sin\Theta \left(1-\tfrac{1}{2}\vartheta^2\right)+\cos\Theta (\vartheta)\right] \times
\nonumber \\
& \left[\sin\Theta (1-\tfrac{1}{2}[a^2\Theta_x^2+\vartheta^2-2a\vartheta \Theta_x])
       +\cos\Theta (a\Theta_x+\tfrac{1}{2}a^2\Theta_{xx}-\vartheta)\right]
\nonumber \\
& \approx \sin^2\!\Theta (-1+\vartheta^2-a\vartheta\Theta_x+\tfrac{1}{2}a^2\Theta_x^2)
+\cos^2\!\Theta (\vartheta^2-a\vartheta\Theta_x)
-\sin\Theta\cos\Theta (a\Theta_x+\tfrac{1}{2}a^2\Theta_{xx}).
\end{align}
Consider instead $n=$odd and $n+1=$even.  This results effectively in $\vartheta\to -\vartheta$.
\be
S_n^z S_{n+1}^z = \sin[-\Theta(x)+\vartheta]\sin[\Theta(x+a)+\vartheta].
\ee
Averaging over these cases will eliminate linear terms in $\vartheta$.  Further, we should average
over right and left displacements, or $a \to -a$.  This will eliminate terms linear in $a$ or first
space derivatives. What results is
\begin{align}
\langle S_n^z S_{n \pm 1}^z \rangle \approx &
-\sin^2\Theta\, (1-\vartheta^2-\tfrac{1}{2}a^2\Theta_x^2) +\cos^2\Theta\, (\vartheta^2)
-\sin\Theta\cos\Theta\, (\tfrac{1}{2}a^2\Theta_{xx})
\nonumber \\
& \approx -\sin^2\Theta\, (1-\tfrac{1}{2}a^2\Theta_x^2) +\vartheta^2
-\sin 2\Theta\, (\tfrac{1}{4}a^2\Theta_{xx})
\end{align}
The term with $\Theta_{xx}$ should be integrated by parts to make it depend on $\Theta_x^2$. Use
\be
\int_{-\infty}^{\infty} dx\ \Theta_{xx} \sin 2\Theta
= \int_{-\infty}^{\infty} dx \left[ \frac{d}{dx} \left(\Theta_x \sin 2\Theta\right)-2\Theta_x^2 \cos 2\Theta \right]
= -\int_{-\infty}^{\infty} dx\ 2\Theta_x^2 \cos 2\Theta .
\ee
Doing this replacement, there are cancellations, resulting in
\be
\label{zz2}
\langle S_n^z S_{n \pm 1}^z \rangle \approx
 -\sin^2\!\Theta +\vartheta^2 +\tfrac{1}{2}a^2\Theta_x^2\, \cos^2\!\Theta .
\ee
It is multiplied by $+D$ in the Hamiltonian.

Next, consider the dipolar part with $x$-components, for $n=$even,
\be
S_n^x S_{n+1}^x = \cos[\Theta(x)+\vartheta] \cos[\Phi(x)+\varphi]
\cos[-\Theta(x+a)+\vartheta] \cos[-\Phi(x+a)+\varphi], \quad n=\text{even}.
\ee
Use expansions keeping up to net cubic order in small terms,
\begin{align}
\cos[\Theta(x)+\vartheta] & \approx \cos\Theta (1-\tfrac{1}{2}\vartheta^2) -\sin\Theta (\vartheta),
\nonumber \\
\cos[\Theta(x+a)-\vartheta] & \approx \cos(\Theta+a\Theta_x+\tfrac{1}{2}a^2\Theta_{xx}-\vartheta)
\nonumber \\
& \approx \cos\Theta \cos(a\Theta_x+\tfrac{1}{2}a^2\Theta_{xx}-\vartheta)
-\sin\Theta \sin(a\Theta_x+\tfrac{1}{2}a^2\Theta_{xx}-\vartheta) 
\nonumber \\
& \approx \cos\Theta[1-\tfrac{1}{2}(a^2\Theta_x^2+\vartheta^2-2a\vartheta\Theta_x)]
-\sin\Theta (a\Theta_x+\tfrac{1}{2}a^2\Theta_{xx}-\vartheta) 
\end{align}
The product of these is
\begin{align}
& \cos[\Theta(x)+\vartheta] \cos[\Theta(x+a)-\vartheta] \approx 
\nonumber \\ 
& \cos^2\!\Theta[1-\vartheta^2+a\vartheta\Theta_x-\tfrac{1}{2}a^2\Theta_x^2]
+\sin^2\!\Theta (a\Theta_x-\vartheta)\vartheta 
-\tfrac{1}{2}\sin 2\Theta \left(a\Theta_x+\tfrac{1}{2}a^2\Theta_{xx}\right).
\end{align}
A similar expression is found for the in-plane angles,
%
\begin{align}
& \cos[\Phi(x)+\varphi] \cos[\Phi(x+a)-\varphi]  \approx 
\nonumber \\
& \cos^2\!\Phi[1-\varphi^2+a\varphi\Phi_x-\tfrac{1}{2}a^2\Phi_x^2]
+\sin^2\!\Phi (a\Phi_x-\varphi)\varphi
-\tfrac{1}{2} \sin 2\Phi \left(a\Phi_x+\tfrac{1}{2}a^2\Phi_{xx} \right).
\end{align}
These four cosines will be multiplied together.  However, we need to also consider the case 
$n=$odd, $n+1=$even, and then average over $n$ being even or odd.  If $n=$odd, effectively the 
signs of the small angles are reversed,
\be
S_n^x S_{n+1}^x = \cos[\Theta(x)-\vartheta] \cos[\Phi(x)-\varphi]
\cos[\Theta(x+a)+\vartheta] \cos[\Phi(x+a)+\varphi], \quad n=\text{odd}.
\ee
The sign reversal applies also to the expanded expressions. One should average over selecting
$n$ even and odd ($\pm\varphi$ and $\pm\vartheta$), which then will cancel out odd powers of 
$\varphi,\vartheta$. At the same time, also average over left and right nearest neighbors 
(i.e., $\pm a$), which cancels out odd powers of $a$, leading to 
\begin{align}
\label{cccc}
\langle S_n^x S_{n \pm 1}^x \rangle \approx &
\cos^2\Theta \cos^2\Phi\, [1-(\vartheta^2+\tfrac{1}{2}a^2\Theta_x^2+\varphi^2+\tfrac{1}{2}a^2\Phi_x^2) ]
+\tfrac{1}{4}\sin 2\Theta \sin 2\Phi\, [a^2\Theta_x \Phi_x]
\nonumber \\
& +\cos^2\Theta \sin^2\Phi\, (-\varphi^2)+\sin^2\Theta \cos^2\Phi (-\vartheta^2)
\nonumber \\
& -\tfrac{1}{4}\cos^2\Theta \sin 2\Phi (a^2\Phi_{xx})-\tfrac{1}{4}\cos^2\Phi \sin 2\Theta (a^2\Theta_{xx}).
\end{align}
The terms in the last line are best converted to squared gradients, by integration by parts in the net
Hamiltonian, such as,  
\begin{align}
\int_{-\infty}^{\infty} dx\ \Phi_{xx} \cos^2\Theta \sin 2\Phi
& = \int_{-\infty}^{\infty} dx\ \left[\frac{d}{dx} \left(\Phi_{x} \cos^2\Theta \sin 2\Phi \right)
+\Phi_x\Theta_x \sin 2\Theta \sin 2\Phi-2\Phi_x^2 \cos^2\Theta \cos 2\Phi\right]
\nonumber \\
& =  \int_{-\infty}^{\infty} dx\ \left[ \Phi_x\Theta_x \sin 2\Theta \sin 2\Phi-2\Phi_x^2 \cos^2\Theta (\cos^2\Phi-\sin^2\Phi)\right].
\end{align}
The other such term has $\Theta$ and $\Phi$ swapped, 
\begin{align}
\int_{-\infty}^{\infty} dx\ \Theta_{xx} \cos^2\Phi \sin 2\Theta
& = \int_{-\infty}^{\infty} dx\ \left[\frac{d}{dx} \left(\Theta_{x} \cos^2\Phi \sin 2\Theta \right)
+\Theta_x\Phi_x \sin 2\Phi \sin 2\Theta-2\Theta_x^2 \cos^2\Phi \cos 2\Theta\right]
\nonumber \\
& =  \int_{-\infty}^{\infty} dx\ \left[ \Theta_x\Phi_x \sin 2\Phi \sin 2\Theta-2\Theta_x^2 \cos^2\Phi (\cos^2\Theta-\sin^2\Theta)\right].
\end{align}
Using these in Eq.\ (\ref{cccc}), after cancellations,  gives the averaged value,
\begin{align}
\langle S_n^x S_{n \pm 1}^x \rangle \approx &
\cos^2\Theta \cos^2\Phi\, (1-\vartheta^2-\varphi^2)
-\tfrac{1}{4}\sin 2\Theta \sin 2\Phi\, (a^2\Theta_x \Phi_x)
\nonumber \\
& -\cos^2\Theta \sin^2\Phi\, (\varphi^2+\tfrac{1}{2}a^2\Phi_x^2)
-\sin^2\Theta \cos^2\Phi\, (\vartheta^2+\tfrac{1}{2}a^2\Theta_x^2).
\end{align}
This is multiplied by $-2D$ to form its contribution to $H$.

The next dipolar part involves
\begin{align}
S_n^y S_{n+1}^y & = \cos[+\Theta(x)+\vartheta] \sin[+\Phi(x)+\varphi]
\cos[-\Theta(x+a)+\vartheta] \sin[-\Phi(x+a)+\varphi], \quad n=\text{even},
\nonumber \\
S_n^y S_{n+1}^y & = \cos[-\Theta(x)+\vartheta] \sin[-\Phi(x)+\varphi]
\cos[+\Theta(x+a)+\vartheta] \sin[+\Phi(x+a)+\varphi], \quad n=\text{odd}.
\end{align}
The product of sine factors for $n=$even are
\begin{align}
& -\sin[\Phi(x)+\varphi]\sin[\Phi(x+a)-\varphi]  \approx
\nonumber \\
& -\sin^2\Phi (1-\varphi^2+a\varphi\Phi_x-\tfrac{1}{2}a^2\Phi_x^2)
-\cos^2\Phi (a\varphi\Phi_x-\varphi^2)
 -\tfrac{1}{2}\sin 2\Phi\, (a\Phi_x+\tfrac{1}{2}a^2\Phi_{xx}).
\end{align}
For $n=$odd, the  $\varphi$ is reversed to $-\varphi$.  This is combined with the same cosine factors of out-of-plane
angles found above.  Then, when averaging over $n=$even and $n=$odd, while also averaging over right and left neighbors,
there results
\begin{align}
\langle S_n^y S_{n \pm 1}^y \rangle \approx &
\cos^2\Theta\sin^2\Phi\, [-1+(\vartheta^2+\tfrac{1}{2}a^2\Theta_x^2+\varphi^2+\tfrac{1}{2}a^2\Phi_x^2)]
+\tfrac{1}{4}\sin 2\Theta \sin 2\Phi\, (a^2\Theta_x\Phi_x)
\nonumber \\
& +\cos^2\Theta \cos^2\Phi (\varphi^2) +\sin^2\Theta \sin^2\Phi (\vartheta^2)
\nonumber \\
& -\tfrac{1}{4} \cos^2\Theta \sin 2\Phi (a^2\Phi_{xx})
+\tfrac{1}{4} \sin^2\Phi \sin 2\Theta (a^2\Theta_{xx}).
\end{align}
Integration by parts can be applied to the last terms, such as already encountered and additionally,
\begin{align}
\int_{-\infty}^{\infty} dx\ \Theta_{xx} \sin^2\Phi \sin 2\Theta
& = \int_{-\infty}^{\infty} dx\ \left[\frac{d}{dx} \left(\Theta_{x} \sin^2\Phi \sin 2\Theta \right)
-\Phi_x\Theta_x \sin 2\Phi \sin 2\Theta-2\Theta_x^2 \sin^2\Phi \cos 2\Theta\right]
\nonumber \\
& =  \int_{-\infty}^{\infty} dx\ \left[-\Phi_x\Theta_x \sin 2\Theta \sin 2\Phi-2\Theta_x^2 \sin^2\Phi (\cos^2\Theta-\sin^2\Theta)\right].
\end{align}
Combining factors and simplifying, one obtains
\begin{align}
\langle S_n^y S_{n \pm 1}^y \rangle \approx &
\cos^2\Theta\sin^2\Phi\, (-1+\vartheta^2+\varphi^2)
-\tfrac{1}{4}\sin 2\Theta \sin 2\Phi\, (a^2\Theta_x\Phi_x)
\nonumber \\
& +\cos^2\Theta \cos^2\Phi\, (\varphi^2+\tfrac{1}{2}a^2\Phi_x^2) +\sin^2\Theta \sin^2\Phi\, (\vartheta^2+\tfrac{1}{2}a^2\Theta_x^2).
\end{align}
In its contribution to $H$, this is multiplied by $+D$. 

The net dipole-dipole interaction in the Hamiltonian is
\be
H_D = D \int \frac{dx}{a}
\left\{ -2\langle S_n^x S_{n \pm 1}^x \rangle +\langle S_n^y S_{n \pm 1}^y \rangle 
+ \langle S_n^z S_{n \pm 1}^z \rangle \right\} = \int \frac{dx}{a} {\cal H}_D.
\ee
There are some cancellations among the terms, and one gets the dipolar energy density,
\begin{align}
\frac{{\cal H}_D}{D} = & -1 -\cos^2\Theta \cos^2\Phi 
+(2 +\cos^2\Phi)\vartheta^2 +(3\cos^2\Theta)\, \varphi^2 
\nonumber \\
& + \cos^2\Theta (2-\cos^2\Phi)\tfrac{1}{2}\Phi_x^2
+(1+\sin^2\Theta\cos^2\Phi) \tfrac{1}{2}\Theta_x^2 
+\tfrac{1}{4} (\sin 2\Theta \sin 2\Phi)\, \Theta_x \Phi_x .
\end{align}
The first line contains (1) a constant and $\Theta,\Phi$ only terms, and (2) quadratic terms in $\varphi,\vartheta$.
The second line contains all the gradient terms.  These all appear in Eq.\ (\ref{H-or}), with a different organization. 

The $K_1$ anisotropy term on an even site depends on 
\begin{align}
\left(S_{\rm evn}^y\right)^2 & = \cos^2\!\theta_{\rm evn} \sin^2\!\phi_{\rm evn}
= \cos^2\!(\Theta+\vartheta) \sin^2\!(\Phi+\varphi) 
\nonumber \\
& \approx \left[(1-\vartheta^2)\cos^2\Theta-\vartheta\sin 2\Theta+\vartheta^2 \sin^2\Theta\right]
\left[(1-\varphi^2)\sin^2\Phi+\varphi\sin 2\Phi+\varphi^2 \cos^2\Phi\right].
\end{align}
For an odd site, this is modified to
\begin{align}
\left(S_{\rm odd}^y\right)^2 & = \cos^2\!\theta_{\rm odd} \sin^2\!\phi_{\rm odd}
= \cos^2\!(-\Theta+\vartheta) \sin^2\!(-\Phi+\varphi)
\nonumber \\
& \approx \left[(1-\vartheta^2)\cos^2\Theta +\vartheta\sin 2\Theta+\vartheta^2 \sin^2\Theta\right]
\left[(1-\varphi^2)\sin^2\Phi - \varphi\sin 2\Phi+\varphi^2 \cos^2\Phi\right].
\end{align}
Taking the average over the sublattices eliminates odd powers of $\varphi,\vartheta$ and results in
\begin{align}
\langle \left(S_n^y\right)^2 \rangle & \approx (1-\vartheta^2-\varphi^2)\cos^2\Theta\sin^2\Phi
+\vartheta^2\sin^2\Theta \sin^2\Phi +\varphi^2\cos^2\Theta \cos^2\Phi -\vartheta\varphi\sin 2\Theta\sin 2\Phi
\nonumber \\
& \approx \cos^2\Theta\sin^2\Phi -\vartheta^2 \cos 2\Theta \sin^2\Phi +\varphi^2\cos^2\Theta \cos 2\Phi
-\vartheta\varphi\sin 2\Theta\sin 2\Phi.
\end{align}
This is multiplied by $-K_1$ to produce the contribution to $H$ in Eq.\ (\ref{H-or}).

For the $K_3$ anisotropy, the calculation is similar and summarized as
\begin{align}
\langle \left(S_n^z\right)^2 \rangle & = \langle \sin^2(\pm \Theta+\vartheta) \rangle
 =  \langle \left(\pm\sin\Theta \cos\vartheta + \cos\Theta\sin\vartheta\right)^2 \rangle
\nonumber \\
& \approx (1-\vartheta^2)\sin^2\Theta+\vartheta^2 \cos^2\Theta
\approx \sin^2\Theta +\vartheta^2 \cos 2\Theta.
\end{align}
This is multiplied by $K_3$ to get the contribution to $H$ in Eq.\ (\ref{H-or}).

For the applied field interaction, one needs to keep up to cubic order terms, so that up to quadratic
terms are retained in the dynamic equations correctly. One has
\begin{align}
S_{\rm evn}^y & =  \cos(\Theta+\vartheta) \sin(\Phi+\varphi)
= (\cos\Theta\cos\vartheta-\sin\Theta\sin\vartheta)(\sin\Phi\cos\varphi+\cos\Phi\sin\varphi)
\nonumber \\
& \approx \left[(1-\tfrac{1}{2}\vartheta^2)\cos\Theta-\left(\vartheta-\tfrac{1}{6}\vartheta^3\right)\sin\Theta\right]
          \left[(1-\tfrac{1}{2}\varphi^2)\sin\Phi+\left(\varphi-\tfrac{1}{6}\varphi^3\right)\cos\Phi\right]
\nonumber \\
& \approx (1-\tfrac{1}{2}\vartheta^2-\tfrac{1}{2}\varphi^2)\cos\Theta\sin\Phi
-\vartheta\varphi\sin\Theta\cos\Phi
\nonumber \\
& -\left(\vartheta-\tfrac{1}{6}\vartheta^3-\tfrac{1}{2}\vartheta\varphi^2\right)\sin\Theta\sin\Phi
  +\left(\varphi-\tfrac{1}{6}\varphi^3-\tfrac{1}{2}\varphi\vartheta^2\right)\cos\Theta\cos\Phi.
\end{align}
On the odd sublattice, reverse signs on $\Theta$ and $\Phi$,
\begin{align}
S_{\rm odd}^y & =  \cos(-\Theta+\vartheta) \sin(-\Phi+\varphi)
= (\cos\Theta\cos\vartheta+\sin\Theta\sin\vartheta)(-\sin\Phi\cos\varphi+\cos\Phi\sin\varphi)
\nonumber \\
& \approx [(1-\tfrac{1}{2}\vartheta^2)\cos\Theta+\left(\vartheta-\tfrac{1}{6}\vartheta^3\right)\sin\Theta]
          [-(1-\tfrac{1}{2}\varphi^2)\sin\Phi+\left(\varphi-\tfrac{1}{6}\varphi^3\right)\cos\Phi]
\nonumber \\
& \approx -(1-\tfrac{1}{2}\vartheta^2-\tfrac{1}{2}\varphi^2)\cos\Theta\sin\Phi
+\vartheta\varphi\sin\Theta\cos\Phi
\nonumber \\
& -\left(\vartheta-\tfrac{1}{6}\vartheta^3-\tfrac{1}{2}\vartheta\varphi^2\right)\sin\Theta\sin\Phi
  +\left(\varphi-\tfrac{1}{6}\varphi^3-\tfrac{1}{2}\varphi\vartheta^2\right)\cos\Theta\cos\Phi .
\end{align}
Then the sublattice average is
\be
\label{SyAB}
\langle S_n^y \rangle \approx  -\left(\vartheta-\tfrac{1}{6}\vartheta^3-\tfrac{1}{2}\vartheta\varphi^2\right)\sin\Theta\sin\Phi
+\left(\varphi-\tfrac{1}{6}\varphi^3-\tfrac{1}{2}\varphi\vartheta^2\right)\cos\Theta\cos\Phi .
\ee
This is multiplied by $-\mu B$ to get the contribution to $H$ in Eq.\ (\ref{H-or}).
That defines all the contributions to $H$ in the opposing rotations assumption, for a general case.

\subsection{The general dynamics}
The discrete dynamics on even-$n$ and odd-$n$ sites each depends on its partial Hamiltonian.  The discrete dynamic 
equations (\ref{HamEq}) show that $S_n^z=\sin\theta_n$ determines the momentum coordinate $z_n$ that is conjugate to 
$\phi_n$,
\be
z_n \equiv \frac{\mu}{\gamma}\sin\theta_n.
\ee
The $n$ index can be dropped and replaced by the sublattice name as we go to the continuum expressions, obtained from
\be
\dot\phi_{\rm evn} = \frac{\partial H_{\rm evn}}{\partial z_{\rm evn}}, \hskip 0.5cm
\dot z_{\rm evn} = -\frac{\partial H_{\rm evn}}{\partial \phi_{\rm evn}}, \hskip 1.0cm
\dot\phi_{\rm odd} = \frac{\partial H_{\rm odd}}{\partial z_{\rm odd}}, \hskip 0.5cm
\dot z_{\rm odd} = -\frac{\partial H_{\rm odd}}{\partial \phi_{\rm odd}}.
\ee
This is equivalent to using twice the full Hamiltonian in place of either $H_{\rm evn}$ or $H_{\rm odd}$, i.e.,
\be
\label{dots-op}
\dot\phi_{\rm evn} = 2\frac{\partial H}{\partial z_{\rm evn}}, \hskip 0.5cm
\dot z_{\rm evn} = -2\frac{\partial H}{\partial \phi_{\rm evn}}, \hskip 1.0cm
\dot\phi_{\rm odd} = 2\frac{\partial H}{\partial z_{\rm odd}}, \hskip 0.5cm
\dot z_{\rm odd} = -2\frac{\partial H}{\partial \phi_{\rm odd}}.
\ee
Using the mapping in (\ref{map}), these will be converted to the continuum angles with
\begin{align}
\theta_{\rm evn} & = +\Theta+\vartheta, \hskip 0.5cm      z_{\rm evn} = \tfrac{\mu}{\gamma}\sin(+\Theta+\vartheta),
\hskip 0.5cm \phi_{\rm evn}=+\Phi+\varphi, \nonumber \\
\theta_{\rm odd} & = -\Theta+\vartheta, \hskip 0.5cm  z_{\rm odd} = \tfrac{\mu}{\gamma}\sin(-\Theta+\vartheta),
\hskip 0.5cm \phi_{\rm odd}=-\Phi+\varphi.
\end{align}
The relations can be inverted as,
\begin{align}
\Theta &= \tfrac{1}{2}(\theta_{\rm evn} -\theta_{\rm odd}), \hskip 1.0cm
\theta_{\rm evn} = \sin^{-1} \tfrac{\gamma}{\mu} z_{\rm evn}, \hskip 0.87cm
\Phi = \tfrac{1}{2}(\phi_{\rm evn}-\phi_{\rm odd}), \nonumber \\
\vartheta &= \tfrac{1}{2}(\theta_{\rm evn} +\theta_{\rm odd}), \hskip 1.0cm
\theta_{\rm odd} = \sin^{-1} \tfrac{\gamma}{\mu} z_{\rm odd}, \hskip 0.8cm
\varphi = \tfrac{1}{2}(\phi_{\rm evn}+\phi_{\rm odd}).
\end{align}

\subsection{Time derivatives of in-plane angles, general ${\cal H}$}
Taking the time derivatives of in-plane angles, while applying the Hamiltonian equations (\ref{dots-op}) produces
\begin{align}
\dot\Phi &= \frac{1}{2}(\dot\phi_{\rm evn}-\dot\phi_{\rm odd})
= \frac{1}{2}\left(2\frac{\partial H}{\partial z_{\rm evn}}-2\frac{\partial H}{\partial z_{\rm odd}}\right)
= \left(\frac{\partial H}{\partial z_{\rm evn}}-\frac{\partial H}{\partial z_{\rm odd}}\right),
\nonumber \\
\dot\varphi &= \frac{1}{2}(\dot\phi_{\rm evn}+\dot\phi_{\rm odd})
= \frac{1}{2}\left(2\frac{\partial H}{\partial z_{\rm evn}}+2\frac{\partial H}{\partial z_{\rm odd}}\right)
= \left(\frac{\partial H}{\partial z_{\rm evn}}+\frac{\partial H}{\partial z_{\rm odd}}\right).
\end{align}
The factors of 2 have all cancelled. Now express as derivative with respect to angles $\Theta$ and $\vartheta$,
doing
\be
\frac{\partial H}{\partial z_{\rm evn}}
= \frac{\partial H}{\partial \Theta}\frac{\partial\Theta}{\partial z_{\rm evn}}
+ \frac{\partial H}{\partial \vartheta}\frac{\partial\vartheta}{\partial z_{\rm evn}},
\hskip 1.0cm
\frac{\partial H}{\partial z_{\rm odd}}
= \frac{\partial H}{\partial \Theta}\frac{\partial\Theta}{\partial z_{\rm odd}}
+ \frac{\partial H}{\partial \vartheta}\frac{\partial\vartheta}{\partial z_{\rm odd}}.
\ee
One needs a simple identity for sines,
\be
\label{dasin1}
\text{if } z = \sin\theta \quad \text{then} \quad
\frac{d z}{d z} = 1 =  \cos\theta\, \frac{d \theta}{d z}
\quad \implies \frac{d \theta}{d z} = \frac{d}{d z} \sin^{-1}z = \frac{1}{\cos\theta}.
\ee
This provides some needed partials,
\begin{align}
\Theta & = \tfrac{1}{2}\left(\sin^{-1} \tfrac{\gamma}{\mu} z_{\rm evn}-\sin^{-1} \tfrac{\gamma}{\mu} z_{\rm odd}\right),
\nonumber \\
& \frac{\partial\Theta}{\partial z_{\rm evn}} = \frac{\tfrac{\gamma}{\mu}}{2\cos(\Theta+\vartheta)},
\hskip 0.9cm
 \frac{\partial\Theta}{\partial z_{\rm odd}} = \frac{-\tfrac{\gamma}{\mu}}{2\cos(-\Theta+\vartheta)} 
= \frac{-\tfrac{\gamma}{\mu} }{2\cos(\Theta-\vartheta)}.
\end{align}
\begin{align}
\vartheta & = \tfrac{1}{2}\left(\sin^{-1} \tfrac{\gamma}{\mu}z_{\rm evn}+\sin^{-1} \tfrac{\gamma}{\mu}z_{\rm odd}\right),
\nonumber \\
& \frac{\partial\vartheta}{\partial z_{\rm evn}} = \frac{\tfrac{\gamma}{\mu}}{2\cos(\Theta+\vartheta)},
\hskip 0.9cm
 \frac{\partial\vartheta}{\partial z_{\rm odd}} = \frac{\tfrac{\gamma}{\mu}}{2\cos(-\Theta+\vartheta)} 
= \frac{\tfrac{\gamma}{\mu}}{2\cos(\Theta-\vartheta)}.
\end{align}
The partials of $H$ with respect to $z$'s are then
\be
\frac{\partial H}{\partial z_{\rm evn}}
= \frac{\tfrac{\gamma}{\mu}}{2\cos(\Theta+\vartheta)}
\left( \frac{\partial H}{\partial \Theta}+\frac{\partial H}{\partial \vartheta} \right),
\hskip 1.0cm
\frac{\partial H}{\partial z_{\rm odd}}
= \frac{\tfrac{\gamma}{\mu}}{2\cos(\Theta-\vartheta)}
\left(-\frac{\partial H}{\partial \Theta}+\frac{\partial H}{\partial \vartheta} \right).
\ee
This leads to unsimplified expressions,
\begin{align}
\frac{\mu}{\gamma} \dot\Phi & = \frac{1}{2\cos(\Theta+\vartheta)}
\left( \frac{\partial H}{\partial \Theta}+\frac{\partial H}{\partial \vartheta} \right)
-\frac{1}{2\cos(\Theta-\vartheta)}
\left(-\frac{\partial H}{\partial \Theta}+\frac{\partial H}{\partial \vartheta} \right),
\nonumber \\
\frac{\mu}{\gamma} \dot\varphi & = \frac{1}{2\cos(\Theta+\vartheta)}
\left( \frac{\partial H}{\partial \Theta}+\frac{\partial H}{\partial \vartheta} \right)
+\frac{1}{2\cos(\Theta-\vartheta)}
\left(-\frac{\partial H}{\partial \Theta}+\frac{\partial H}{\partial \vartheta} \right).
\end{align}
Using the fact that $\vartheta$ is a small angle, and generally we may want up to quadratic terms in it,
requires expansions,
\begin{align}
\frac{1}{\cos(\Theta \pm \vartheta)} & \approx \frac{1}{(1-\tfrac{1}{2}\vartheta^2)\cos\Theta \mp \vartheta \sin\Theta}
= \frac{1}{\cos\Theta(1 \mp \vartheta\tan\Theta-\frac{1}{2}\vartheta^2)}
\nonumber \\
& \approx \sec\Theta\left[1 \pm \vartheta \tan\Theta+\tfrac{1}{2}\vartheta^2+\vartheta^2\tan^2\!\Theta\right]
\approx \sec\Theta\left[1 \pm \vartheta \tan\Theta+\vartheta^2\left(\tan^2\!\Theta+\tfrac{1}{2}\right)\right].
\end{align}
This produces the needed combinations,
\begin{align}
\label{combos}
\frac{1}{\cos(\Theta+\vartheta)}+\frac{1}{\cos(\Theta-\vartheta)} & \approx
2\sec\Theta
\left[1 +\vartheta^2\left(\tan^2\!\Theta+\tfrac{1}{2}\right)\right], \nonumber \\
\frac{1}{\cos(\Theta+\vartheta)}-\frac{1}{\cos(\Theta-\vartheta)} & \approx
2\vartheta\sec\Theta\tan\Theta.
\end{align}
Finally that gives the time derivatives of the in-plane continuum angles in the opposing rotations assumption,
\begin{align}
\label{dot-in-orA}
\frac{\mu}{\gamma} \dot\Phi & =
\sec\Theta \left\{ \left[1 +\vartheta^2\left(\tan^2\!\Theta+\tfrac{1}{2}\right)\right]
\frac{\partial H}{\partial \Theta}
+\vartheta \tan\Theta \frac{\partial H}{\partial \vartheta} \right\},
\nonumber \\
\frac{\mu}{\gamma} \dot\varphi & =
\sec\Theta \left\{ \vartheta\tan\Theta\frac{\partial H}{\partial \Theta}
+\left[1 +\vartheta^2\left(\tan^2\!\Theta+\tfrac{1}{2}\right)\right]
\frac{\partial H}{\partial \vartheta} \right\}.
\end{align}
In real application, the $\vartheta^2$ term in (1st order) $\dot\Phi$ can likely be dropped,
In contrast, $\dot\varphi$ is 2nd order small and so the $\vartheta^2$ term should be kept.

The partial derivatives of $H$ really mean functional derivatives, for example,
\be
\frac{\partial H}{\partial \Theta} \longrightarrow \frac{\delta \cal H}{\delta \Theta}
= -\frac{d}{dx} \left(\frac{\partial \cal H}{\partial \Theta_x}\right)+\frac{\partial \cal H}{\partial \Theta}.
\ee
Additionally, ${\cal H}$ should be expanded to 3rd order small terms before finding its derivatives,
if the $\dot\varphi$ equation is to include all 2nd order small terms.  This is necessary for
the Hamiltonian method to agree exactly with the expansion of the discrete equations,  especially for $\dot\varphi$
and $\dot\vartheta$.

After a tedious calculation, the equations (\ref{dot-in-orA}) resulting from the Hamiltonian in Eq.\ (\ref{H-or}) 
do indeed reproduce the dynamic equations (\ref{dPhidt0}) and (\ref{dphidt0}) found more directly by expansion of the 
discrete torque equations of motion.

\subsection{Time derivatives of out-of-plane angles, general ${\cal H}$}
Consider these time derivatives for opposing rotations,
\begin{align} 
\dot\Theta &= \tfrac{1}{2}(\dot\theta_{\rm evn} -\dot\theta_{\rm odd}) =
\tfrac{1}{2} \tfrac{d}{dt} \left(\sin^{-1} \tfrac{\gamma}{\mu} z_{\rm evn}-\sin^{-1} \tfrac{\gamma}{\mu} z_{\rm odd} \right),
\nonumber \\
\dot\vartheta &= \tfrac{1}{2}(\dot\theta_{\rm evn} +\dot\theta_{\rm odd}) =
\tfrac{1}{2} \tfrac{d}{dt} \left(\sin^{-1} \tfrac{\gamma}{\mu} z_{\rm evn}+\sin^{-1} \tfrac{\gamma}{\mu} z_{\rm odd} \right).
\end{align}
Using the identity (\ref{dasin1}) and the chain rule gives 
\begin{align}
\dot\Theta &= \frac{\gamma}{2\mu} \left[\frac{\dot z_{\rm evn}}{\cos(\Theta+\vartheta)}
-\frac{\dot z_{\rm odd}}{\cos(-\Theta+\vartheta)}\right]
=  \frac{\gamma}{2\mu} \left[\frac{\dot z_{\rm evn}}{\cos(\Theta+\vartheta)}
-\frac{\dot z_{\rm odd}}{\cos(\Theta-\vartheta)}\right], 
\nonumber \\
\dot\vartheta &= \frac{\gamma}{2\mu} \left[\frac{\dot z_{\rm evn}}{\cos(\Theta+\vartheta)}
+\frac{\dot z_{\rm odd}}{\cos(-\Theta+\vartheta)}\right]
=  \frac{\gamma}{2\mu} \left[\frac{\dot z_{\rm evn}}{\cos(\Theta+\vartheta)}
+\frac{\dot z_{\rm odd}}{\cos(\Theta-\vartheta)}\right]. 
\end{align}
Insert time derivatives as found from the averaged $H$, Eq.\ (\ref{dots-op}) (the factors of 2 cancel),
\begin{align}
\frac{\mu}{\gamma} \dot\Theta &=
-\frac{1}{\cos(\Theta+\vartheta)} \frac{\partial H}{\partial \phi_{\rm evn}}
+\frac{1}{\cos(\Theta-\vartheta)} \frac{\partial H}{\partial \phi_{\rm odd}}. 
\nonumber \\
\frac{\mu}{\gamma} \dot\vartheta &=
-\frac{1}{\cos(\Theta+\vartheta)} \frac{\partial H}{\partial \phi_{\rm evn}}
-\frac{1}{\cos(\Theta-\vartheta)} \frac{\partial H}{\partial \phi_{\rm odd}}. 
\end{align}
We want derivatives with respect to  $\Phi$ and $\varphi$, which are in the continuum $H$, obtained via
\be
\frac{\partial H}{\partial \phi_{\rm evn}}
= \frac{\partial H}{\partial \Phi}\frac{\partial\Phi}{\partial \phi_{\rm evn}}
+ \frac{\partial H}{\partial \varphi}\frac{\partial\varphi}{\partial \phi_{\rm evn}},
\hskip 1.0cm
\frac{\partial H}{\partial \phi_{\rm odd}}
= \frac{\partial H}{\partial \Phi}\frac{\partial\Phi}{\partial \phi_{\rm odd}}
+ \frac{\partial H}{\partial \varphi}\frac{\partial\varphi}{\partial \phi_{\rm odd}}.
\ee
It is easy to get the required derivatives:
\begin{align}
\Phi &=\tfrac{1}{2}(\phi_{\rm evn}-\phi_{\rm odd}), \hskip 0.8cm
\frac{\partial \Phi}{\partial \phi_{\rm evn}} =\tfrac{1}{2}, \hskip 0.8cm
\frac{\partial \Phi}{\partial \phi_{\rm odd}} =-\tfrac{1}{2}. 
\nonumber \\
\varphi &=\tfrac{1}{2}(\phi_{\rm evn}+\phi_{\rm odd}), \hskip 0.8cm
\frac{\partial \varphi}{\partial \phi_{\rm evn}} =\tfrac{1}{2}, \hskip 0.8cm
\frac{\partial \varphi}{\partial \phi_{\rm odd}} =+\tfrac{1}{2}. 
\end{align}
Putting these together,
\be
\frac{\partial H}{\partial \phi_{\rm evn}}
= \frac{1}{2} \left( \frac{\partial H}{\partial \Phi}+\frac{\partial H}{\partial \varphi} \right),
\hskip 1.0cm
\frac{\partial H}{\partial \phi_{\rm odd}}
= \frac{1}{2} \left( -\frac{\partial H}{\partial \Phi} +\frac{\partial H}{\partial \varphi} \right).
\ee
This gives unsimplified time derivatives, 
\begin{align}
\frac{\mu}{\gamma} \dot\Theta &=
-\frac{1}{\cos(\Theta+\vartheta)}
\frac{1}{2} \left( \frac{\partial H}{\partial \Phi}+\frac{\partial H}{\partial \varphi} \right)
+\frac{1}{\cos(\Theta-\vartheta)}
\frac{1}{2} \left( -\frac{\partial H}{\partial \Phi} +\frac{\partial H}{\partial \varphi} \right)
\nonumber \\
\frac{\mu}{\gamma} \dot\vartheta &=
-\frac{1}{\cos(\Theta+\vartheta)}
\frac{1}{2} \left( \frac{\partial H}{\partial \Phi}+\frac{\partial H}{\partial \varphi} \right)
-\frac{1}{\cos(\Theta-\vartheta)}
\frac{1}{2} \left( -\frac{\partial H}{\partial \Phi} +\frac{\partial H}{\partial \varphi} \right)
\end{align}
Using the combinations in (\ref{combos}), this leads to the time derivatives of the out-of-plane
angles in the opposing rotations scenario,
\begin{align}
\label{dot-out-orA}
\frac{\mu}{\gamma} \dot\Theta & = -\sec\Theta \left\{
\left[1 +\vartheta^2\left(\tan^2\!\Theta+\tfrac{1}{2}\right)\right] \frac{\partial H}{\partial\Phi}
+\vartheta\tan\Theta \frac{\partial H}{\partial \varphi} \right\}.
\nonumber \\
\frac{\mu}{\gamma} \dot\vartheta &= -\sec\Theta
\left\{ \vartheta\tan\Theta \frac{\partial H}{\partial\Phi}
+\left[1 +\vartheta^2\left(\tan^2\!\Theta+\tfrac{1}{2}\right)\right] \frac{\partial H}{\partial \varphi} \right\},
\end{align}
These involve a certain symmetry compared to the in-plane angle equations. It is likely that the $\vartheta^2$ term 
in $\dot\Theta$ can be dropped, as $\Theta$ is considered to be large. In the equation for $\dot\vartheta$, 
the $\vartheta^2$ term should be kept.

After further tedious calculations, the equations (\ref{dot-out-orA}) obtained with the Hamiltonian in Eq.\ (\ref{H-or}) 
do indeed reproduce the dynamic equations (\ref{dThetadt0}) and (\ref{dthetadt0}) found by expansion of the
discrete torque equations of motion.

\end{widetext}


\begin{thebibliography}{99}

\bibitem{Wang06} Wang R F, Nisoli C, Freitas R S, Li J, McConville W, Cooley B J,  Lund M S, Samarth N,
        Leighton C,  Crespi V H and Schiffer P,
        Artificial spin ice in a geometrically frustrated lattice of nanoscale ferromagnetic islands,
        {Nature} \textbf{439} 303--306 (2006).
        \url{https://doi.org/10.1038/nature04447}

\bibitem{Nisoli13} Nisoli C, Moessner R and Schiffer P,
        Colloquium: Artificial spin ice: Designing and imaging magnetic frustration,
        {Rev. Mod. Phys.} \textbf{85} 1473 (2013).
        \url{https://doi.org/10.1103/RevModPhys.85.1473}

\bibitem{skjaervo19}  Skj\ae rv\o\ S H, Marrows C H,  Stamps R L  and Heyderman L J,
        Advances in artificial spin ice,
        {Nat. Rev. Phys.} \textbf{2} 13--28 (2020).
	\url{https://doi.org/10.1038/s42254-019-0118-3}

\bibitem{Ostman18} \"Ostman E, Arnalds U B, Kapaklis V, Taroni A and Hj\o varsson B,
        Ising-like behaviour of mesoscopic magnetic chains,
        {J. Phys.: Condens. Matt.} \textbf{30} 365301 (2018).
	\url{https://doi.org/10.1088/1361-648X/aad0c1}

\bibitem{Cisternas21} Cisternas J, \textit{et al.},
        Stable and unstable trajectories in a dipolar chain,
        {Phys. Rev. B} \textbf{103} 134443 (2021).
	\url{https://doi.org/10.1103/PhysRevB.103.134443}

\bibitem{Anand21} Anand M,
        Thermal and dipolar interaction effect on the relaxation in a linear chain of magnetic nanoparticles,
        J. Magn. Magn. Mater. \textbf{522} 167538 (2021).
	\url{https://doi.org/10.1016/j.jmmm.2020.167538}

\bibitem{Wysin24} Wysin G M, 
	Transverse magnetic field effects on metastable states of magnetic island chains,
	J. Magn. Magn. Mater. \textbf{610}, 172582 (2024).
	\url{https://doi.org/10.1016/j.jmmm.2024.172582}

\bibitem{Wysin26} Wysin G M, 
	Domain walls in a dipole-coupled transverse magnetic island chain,
	Magnetism \textbf{2026}, Vol.\ 6, Issue 2, Art.\ 18, 
	\url{https://doi.org/10.3390/magnetism6020018}

\bibitem{Mik80} Mikeska H J, 
	Nonlinear dynamics of classical one-dimensional antiferromagnets,
	J. Phys. C: Solid State Phys. \textbf{13}, 1913 (1980).
	\url{https://doi.org/10.1088/0022-3719/13/15/015}

\bibitem{FlugMik83} Fl\"uggen N, and Mikeska H J, 
	On the nonlinear dynamics of the easy-plane antiferromagnetic chain in an external field,
	Solid State Commun. \textbf{48}, 293 (1983).
	\url{https://doi.org/10.1016/0038-1098(83)90290-9}

\bibitem{GalIv18} Galkina E G, and Ivanov B A,
	Dynamic solitons in antiferromagnets,
	Low Temp. Phys. \textbf{44}, 618 (2018).
	\url{https://doi.org/10.1063/1.5041427}

\bibitem{Heisenberg28} Heisenberg W, 1928 
	Zur Theorie des Ferromagnetismus (On the theory of ferromagnetism).
	Z. Physik \textbf{49} 619--636 (1928).
	\url{https://doi.org/10.1007/BF01328601}

\bibitem{Jiles91} Jiles David,
	 \textit{Introduction to Magnetism and Magnetic Materials}, 
	Ch 11 (Chapman and Hall, London and New York, 1st ed., 1991).

\bibitem{Pires+89} Pires A S T, Talim S L, and Costa B V, 
	Solitons in one-dimensional antiferromagnetic chains, 
	Phys. Rev. B \textbf{39}, 7149 (1989).
	\url{https://doi.org/10.1103/PhysRevB.39.7149}

\bibitem{Gerling+86} Gerling R W, Standinger M, and Landau D P, 
	Solitons in the classical xy-chain,
	J. Magn. Mater. \textbf{54}, 819 (1986).
	\url{https://doi.org/10.1016/0304-8853(86)90268-4}
	
\bibitem{sGreview} Caudrey P J, Eilbeck J C, and Gibbon John D,
	The sine-Gordon equation as a model classical field theory,
	Nuov Cim B \textbf{25}, 497 (1975).
	\url{https://doi.org/10.1007/BF02724733}
	

\end{thebibliography}
\end{document}